\documentclass[]{article}
\usepackage{proceed2e}

\usepackage{times}
\usepackage{graphicx}
\usepackage{amssymb}
\usepackage{mathtools}
\usepackage{amsmath}
\usepackage{caption}
\usepackage{subcaption}
\newcommand{\Real}{\mbox{Re}}
\newcommand{\Imag}{\mbox{Im}}

\newcommand{\ci}{\perp\!\!\!\perp}
\newcommand{\BF}[1]{\mathbf{#1}}

\usepackage{color}

\title{Bayesian Structure Learning for Stationary Time Series}

\author{Alex Tank \\ University of Washington \\ alextank@uw.edu \And Nicholas J.  Foti \\ University of Washington \\ nfoti@uw.edu \And Emily B. Fox \\ University of Washington \\ ebfox@uw.edu}  
\begin{document}

\maketitle
\begin{abstract}
While much work has explored probabilistic graphical models for independent data, less attention has been paid to time series. The goal in this setting is to determine conditional independence relations between entire time series, which for stationary series, are encoded by zeros in the inverse spectral density matrix. We take a Bayesian approach to structure learning, placing priors on (i) the graph structure and (ii) spectral matrices given the graph.  We leverage a Whittle likelihood approximation and define a conjugate prior---the \emph{hyper complex inverse Wishart}---on the complex-valued and graph-constrained spectral matrices. Due to conjugacy, we can analytically marginalize the spectral matrices and obtain a closed-form marginal likelihood of the time series given a graph. Importantly, our analytic marginal likelihood allows us to avoid inference of the complex spectral matrices themselves and places us back into the framework of standard (Bayesian) structure learning.  In particular, combining this marginal likelihood with our graph prior leads to efficient inference of the time series graph itself, which we base on a stochastic search procedure, though any standard approach can be straightforwardly modified to our time series case. We demonstrate our methods on analyzing stock data and neuroimaging data of brain activity during various auditory tasks.

\end{abstract}

\section{INTRODUCTION}
Probabilistic graphical models (PGMs)---which compactly encode a set of conditional independence statements
---have become a defacto tool for defining probabilistic models over large sets of random variables.
When faced with time series, dynamic Bayesian networks (DBNs) are commonly
deployed and specify sparse between- and within-time dependencies, often
encoded by a \emph{template model} replicated across time to straightforwardly
model the growing set of random variables~\cite{Koller:2009}.  Learning
template models requires specifying the set of dependency lags to be considered
\cite{Ziebart:2007,Siracusa:2009}. In many applications, one instead aims to
infer conditional independence between entire data streams accounting for
interactions at all possible lags, represented by a \emph{time series graphical
model} (TGM).  For example, imagine recording brain activity from multiple
regions of the brain over time. Inference of a TGM in this setting would
provide insight into the functional connectivity of different brain regions, an
object of substantial scientific interest \cite{Sporns:2010, Medkour:2010}.
TGMs have also been applied to intensive care monitoring \cite{Gather:2000} and
financial time series \cite{Songsri:2011}.

The pioneering work of Dahlhaus \cite{Dahlhaus:2000} introduced the concept of undirected graphical models for stationary time series. The key insight was to transform the series to the \emph{frequency domain} and express the graph relationships in the resulting spectral representation. For jointly Gaussian stationary time series, Dahlhaus \cite{Dahlhaus:2000} showed that conditional independencies between time series are encoded by zeros in the inverse spectral density matrices. This result is the frequency-domain analog to Gaussian graphical modeling in the i.i.d. (non-time-series) setting, where zeros in the inverse covariance matrix, or \emph{precision matrix}, encode the conditional independencies between observed dimensions \cite{Lauritzen:1996}.  Dahlhaus' insight was first exploited to perform independent hypothesis tests of conditional independence between each pair of time series \cite{Dahlhaus:2000}, with more recent work correcting for multiple comparisons \cite{Matsuda:2006, Wolstenholme:2015}.

A likelihood-based approach leveraging the \emph{Whittle approximation} \cite{Whittle:1953} has also been introduced \cite{Bach:2004}. The Whittle approximation casts the likelihood in the frequency domain with terms depending on the spectral density matrices critical to TGM structure learning, and independently so across frequencies. One approach scores graphs using AIC \cite{Bach:2004}. A recent penalized likelihood variant \cite{Jung:2014} places a joint graphical lasso \cite{Danaher:2014} across frequencies to enforce a common zero pattern in the spectral density matrices. A penalized likelihood approach restricted to finite vector autoregressive processes has also been considered \cite{Songsri:2011}. 
 
We instead consider a Bayesian approach to TGM structure learning, with all the benefits garnered from the Bayesian paradigm, including modeling within a generative framework where information from multiple sources can integrated and combined with available prior knowledge.  For example, neural data are notoriously noisy, and robust inferences often rely on integrating time series across multiple trials and individuals or recording platforms (e.g., EEG/MEG).  Our approach also leverages the Whittle likelihood.  We then introduce a novel hyper Markov law \cite{Dawid:1993}, the \emph{hyper complex inverse Wishart} distribution, that serves as a conjugate prior for the spectral density matrices whose inverses have a zero pattern specified by a graph. For decomposable graphs, this formulation leads to a closed-form expression for the marginal likelihood of a multivariate time series given a graph. By placing a prior on graph structures, we achieve a fully Bayesian approach to TGM structure learning for stationary time series. For our graph prior, we consider a multiplicity correcting prior \cite{Carvalho:2009}.  Our analytic expression for the marginal likelihood is critical to the practicality of our approach since we can avoid inference of the large set of high-dimensional, complex spectral density matrices.  In particular, for a length $T$ series of dimension $p$, there are $T$ $p\times p$ spectral matrices to consider.  In the i.i.d. setting, inference of just a single $p\times p$ graph-constrained covariance matrix is challenging; in this setting, inference of the $T$ $p \times p$ matrices is prohibitive.
 
Hyper Markov laws based on the hyper inverse Wishart are a popular tool for Bayesian graphical model selection in the i.i.d. setting \cite{Giudici:1999, jones:2005}. Indeed, many powerful Bayesian structure learning algorithms based on this framework have been developed, both for decomposable \cite{Scott:2008, Green:2013} and non-decomposable \cite{Moghaddam:2009, mohammadi2015} graphs. By framing TGM structure learning in this common framework, we are able to apply existing state-of-the-art inference machinery for standard structure learning to the time series case. In this paper we use the feature-inclusion stochastic search (FINCS) procedure \cite{Scott:2008} for inference in decomposable models; however, many other MCMC and search schemes may be used.  Importantly, future computational advances in Bayesian inference for i.i.d. graphical models may be easily extended using our framework to the time series case.

We test our methods on data simulated from vector autoregressive models with randomly generated TGMs. Our approach reaches almost perfect TGM recovery as the length of the time series or number of independent replicates increases. We then demonstrate the utility of our methods on a global stock indices dataset and MEG neuroimaging data of auditory attention switching tasks.  In both cases we find meaningful, intuitive structure in the data.

Our paper is organized as follows.  We provide background on graphical models and stationary time series in Sec.~\ref{sec:background}. Our proposed TGM method is in Sec.~\ref{sec:method}, first introduced in the context of multiple independent realizations and then adapted to perform efficient inference of the TGM from only a single realization. In Sec.~\ref{sec:inference}, we discuss how existing Bayesian structure learning methods may be modified to fit our formulation.  Simulated results are in Sec.~\ref{sec:sim}, with our stock and MEG analyses in Secs.~\ref{sec:stock} and \ref{sec:meg}, respectively.
%

\section{BACKGROUND} \label{sec:background}
\vspace{-0.05in}
\subsection{Graphs}
Let $G = (V,E)$ be an undirected graph with vertex set $V = \{1, \ldots, p\}$ and edge set $E$, where $E \subset \{(i,j) \in V \times V: i\neq j\}$. Nodes $i$ and $j$ are adjacent, or \emph{neighbors}, if $(i,j) \in E$. A \emph{complete graph} is one having $(i,j) \in E$ for every $i,j \in V$ and complete subgraphs $C \subset V$ are termed \emph{cliques}. A triple of subgraphs $(A,S,B)$ where $V = A \cup B$ and $S = A \cap B$ with $S$ complete is called a \emph{decomposition} if every path from a node in $A$ to a node in $B$ must pass through $S$, the \emph{separator}.  Recursively decomposing $A$ and $B$ in this fashion results in the \emph{prime components} of a graph. If the prime components are complete then the graph is \emph{decomposable}. We let the sets ${\mathcal{C}} = \{C_1, \ldots , C_K \}$ and ${\mathcal{S}} = \{S_2, \ldots, S_K \}$ each denote the prime components and their separators, respectively,  generated by the decomposition. For simplicity, we restrict our attention to decomposable graphs but stress that our formulation is extensible to the non-decomposable case (see Sec.~\ref{sec:discussion}).

\subsection{Hyper Markov distributions}
\label{sec:hyperMarkov}
For a given set of of random variables $X$, with realization $x \in \mathcal{X}$, dimensionality $p$, and joint density $p(x)$, an undirected graphical model $G$ can be constructed by stating that an edge $(i,j) \notin E$ if $X_i$ and $X_j$ are conditionally independent given the remaining variables, i.e. $X_j  \ci X_i | X_{Z_{ij}}$ where $Z_{ij} = V \setminus \{i,j\}$. If the graph is decomposable, the joint density decomposes over cliques and separators:
\begin{equation}
p(x) = \frac{\prod_{C \in \mathcal{C}} p(x_C)}{\prod_{S \in \mathcal{S}} p(x_S)}
\end{equation}
where $p(x_A)$ for $A \subset V$ denotes the marginal distribution of the set of variables $x_A$. 

A hyper Markov law \cite{Dawid:1993} is a distribution over probability measures that is concentrated on distributions that obey the Markov properties specified by $G$.  
Examples include the hyper Wishart and hyper Dirichlet distribution \cite{Dawid:1993, Giudici:1999}. Such distributions have proven pivotal in Bayesian graphical modeling by serving as conjugate priors for the graph parameters conditioned on the graph structure $G$.  For example, in Gaussian graphical models (GGMs), the hyper inverse Wishart distribution provides a conjugate prior for covariance matrices that obey a zero pattern in the precision, as specified by $G$. By integrating over the hyper Markov distribution, one can obtain the \emph{marginal likelihood} of the data conditioned on the structure $G$ alone. 
 
 \subsection{Stationary time series}\vspace{-.1in}
Let $X(t) = (X_1(t),..., X_p(t))^T \in \mathbb{R}^{p}$ for $t \in \mathbb{Z}$ be a multivariate Gaussian stationary time series such that:\vspace{-.05in}
\begin{align}
E(X(t)) &= \mu \,\,\,\,\,\,\ \mbox{   $\forall t \in \mathbb{Z}$ } \\
\mbox{Cov}(X(t), X(t + h)) &= \Gamma(h) \,\,\,\,\,\, \mbox{           $\forall t, h \in \mathbb{Z}$}. 
\end{align}
A time series probabilistic graphical model (TGM), $G = (V, E)$, may be constructed by letting $(i,j) \notin E$ denote that the entire time series $X_i(:)$ and $X_j(:)$ are conditionally independent given the remaining collection of time series $X_{Z_{ij}}$ where $Z_{ij} = V \setminus \{i, j\}$. For the Gaussian stationary series we consider, one can show that conditional independence holds between time series iff \cite{Dahlhaus:2000} 
\begin{align}
\text{Cov}(X_i(t), X_j(t + h)|X_{Z_{ij}}) = 0 \,\,\,\, \forall h \in \mathbb{Z}. 
\end{align}
 The \emph{spectral density matrix} of a stationary time series is defined as the Fourier transform of the lagged covariance matrices, $\Gamma(h) = \text{Cov}(X(t), X(t +h))$:
 \begin{equation}\label{eq:spec} \vspace{-.1in}
 S(\lambda) = \sum_{h = - \infty}^{\infty}\Gamma(h) e^{-i \lambda h}
 \end{equation}
 for $\lambda \in [0, 2\pi]$ and $S(\lambda) \in \mathbb{C}^{p \times p}$ and Hermitian positive definite. 
The marginal dependencies between time series are captured by $S(\lambda)$, and from Eq.~\eqref{eq:spec}, $S(\lambda)_{ij} = 0$ for all $\lambda \in [0, 2 \pi]$ iff $\Gamma(h)_{ij} = 0$ for all $h \in \mathbb{Z}$. Furthermore, conditional independence between Gaussian stationary time series holds iff 
\begin{equation} \label{eq:inv_zero}
S(\lambda)^{-1}_{ij} = 0 \,\,\,\,\, \forall \lambda \in [0, 2 \pi] ,
\end{equation}
implying that inferring zeros in the inverse spectral density matrices across frequencies equates with inferring the TGM structure \cite{Dahlhaus:2000}. More background on the spectral approach to time series is presented in the Supplement.

\section{A BAYESIAN APPROACH}\label{sec:method}
There are two standard approaches to Bayesian inference in graphical models: (1) placing a prior that jointly specifies the graph structure and associated parameters or (2) placing a prior on graph structures and then a prior on parameters given a graph; both rely on specifying a likelihood model.  We opt for the second approach and describe the various components in this section.  At a high level, our methods combine existing Whittle likelihood based methods \cite{Bach:2004, Jung:2014} with the hyper Markov framework to Bayesian graphical modeling \cite{jones:2005, Giudici:1999}.  In the context of our TGMs, we introduce a conjugate \emph{hyper complex inverse Wishart} prior on graph-constrained spectral density matrices.  By integrating out the spectral density matrices, we obtain a marginal likelihood of the time series given the graph structure, $G$, allowing us to straightforwardly leverage state-of-the-art computational methods for i.i.d. Bayesian structure learning. 

\subsection{Whittle likelihood}
Let $\BF{X} = [X(1), \ldots, X(T)]$, with $x(t) \in \mathbb{R}^p$ a realization of a $p$-dimensional stationary Gaussian time series observed at $T$ time points, and $\BF{X}_{1:N} = \{\BF{X}^1, \ldots, \BF{X}^N\}$ be the collection of $N$ independent realizations.  We move to the frequency domain by transforming each $\BF{X}^i$ using a discrete Fourier transform.  Let $d_{nk} \in \mathbb{C}^p$ denote the discrete Fourier coefficient associated with the $n$th time series at frequency $\lambda_k = \frac{2 \pi k}{T}$:
\begin{align}
d_{nk} = \frac{1}{T} \sum_{t = 0}^{T - 1} x_n(t) e^{- i \lambda_k t }.
\end{align}
The Whittle approximation \cite{Whittle:1953} assumes the Fourier coefficients are independent \emph{complex normal random variables} with mean zero and covariance given by the corresponding spectral density matrix $S_k = S(\lambda_k)$:
\begin{align}
	d_{nk} \sim \mathcal{N}_c(0,S_k) \quad k=0,\dots,T-1,
\end{align}
such that the likelihood of $\BF{X}_{1:N}$ is approximated as
\begin{align} \label{eq:whittle}
p(\BF{X}_{1:N}|S_{0:T-1}) \approx \prod_{n = 1}^{N} \prod_{k = 0}^{T - 1} \frac{1}{\pi^p |S_k|} e^{-d_{nk}^* S^{-1}_k d_{nk}},
\end{align}
where $\frac{1}{\pi^p |S|} e^{-z^* S^{-1} z}$ is the density of a complex normal distribution, $\mathcal{N}_c(0,S)$, with $S \in \mathbb{C}^{p\times p}$ and Hermitian positive definite.  See the Supplement.  The Whittle approximation holds asymptotically with large $T$ \cite{Brockwell:1991,Brillinger:2001,Whittle:1953}.  This approximation has been used in the Bayesian context in \cite{Rosen:2007,Krafty:2015} 

Recall that conditional independencies are encoded in the off diagonal elements of $S_k^{-1}$. If time series $X_i(t)$ and $X_j(t)$ are conditionally independent, then the Whittle approximation says that as $T$ gets large the $i$th and $j$th elements of the Fourier coefficients $d_{nk}$ are conditional independent across all frequencies. Thus, if $G$ is decomposable, Eq.~\eqref{eq:whittle} can be rewritten as
\begin{align}
	\label{eq:whittlegraph}
p(\BF{X}_{1:N}| G, S_{0:(T - 1)})& \approx \\
\,\,\,\,\,\,\,\,\,\, \prod_{k = 0}^{T - 1} &\frac{\prod_{C \in \mathcal{C}}  \frac{1}{\pi^{N |C|} |S_{kC}|^N} e^{-\text{tr} P_{kC} S_{kC}^{-1}}}{ \prod_{S \in \mathcal{S}}  \frac{1}{\pi^{N |S| }|S_{kS}|^N} e^{-\text{tr} P_{kS} S_{kS}^{-1}}} \nonumber
\end{align}
where 
\begin{align}
	\label{eq:periodogram}
	P_k = \sum_{n = 1}^{N} d_{nk} d_{nk}^*
\end{align}
is the aggregate \emph{periodogram} over the $N$ time series at frequency $\frac{2 \pi k}{T}$. For $A \subset V$, $S_{kA}$ and $P_{kA}$ are the restriction of both matrices to the elements in $A$ and $|A|$ denotes the cardinality of the set $A$.

\subsection{Hyper complex inverse Wishart prior on graph-constrained spectral density matrices}
We seek a prior for the spectral density matrices, $S_k$, whose inverses each have zeros dictated by a graph $G$. Recall that these $S_k$ matrices are complex-valued and restricted to be Hermitian positive definite.  As discussed in Sec.~\ref{sec:hyperMarkov}, the hyper inverse Wishart distribution serves as a prior for real-valued, positive-definite matrices with pre-specified zeros in the inverse, and is a conjugate prior for the covariance of a zero-mean GGM.  Motivated by the connection between GGMs and our TGMs, and the analogous structure of our TGM-based Whittle likelihood of Eq.~\eqref{eq:whittlegraph} to that of a GGM with $N$ i.i.d. observations, we propose a novel \emph{hyper complex inverse Wishart} prior with density function
\begin{align} 
p(\Sigma|\delta,W,G ) &= \propto {\BF 1}_{\Sigma \in M^{+}(G)} |\Sigma|^{-(\delta + 2 p)} e^{- \text{tr} W \Sigma^{-1}}
\end{align}
for \emph{degrees of freedom} $\delta > 0$, \emph{scale matrix} $W \in \mathbb{C}^{p \times p} $ positive definite and Hermitian, and graph $G$.  We have used an analogous parameterization to that of the hyper inverse Wishart \cite{Dawid:1993}. Here, $\Sigma \in M^{+}(G)$ denotes that $\Sigma$ is in the set of all Hermitian positive-definite matrices with $\left(\Sigma^{-1}\right)_{ij}=0$ for all $(i,j) \notin E$.  When $G$ is decomposable, the normalization constant is available and the density decomposes over cliques and separators:
\begin{align}
p(\Sigma|\delta,W,G ) &= \frac{\prod_{C \in \mathcal{C}} \mbox{IW}_c(\Sigma_C|\delta,W_C)}{\prod_{S \in \mathcal{S}} \mbox{IW}_c(\Sigma_C|\delta,W_C)}\\
&= \frac{\prod_{C \in \mathcal{C}} B(W_C, \delta) |\Sigma_{C}|^{-(\delta + 2 |C|)} e^{-\text{tr} W_C \Sigma_{C}^{-1}}} {\prod_{S \in \mathcal{S}}B(W_S, \delta) |\Sigma_{S}|^{-(\delta + 2 |S|)} e^{-\text{tr} W_S \Sigma_{S}^{-1}}},
\end{align}
where $\mbox{IW}_c$ denotes the complex inverse Wishart \cite{Brillinger:2001} detailed in the Supplement with normalizer 
\begin{align}
B(W,\delta) = \frac{|W|^{\delta + p}}{\pi^{\frac{p(p - 1)}{2}} \prod_{j = 1}^{p} (\delta + p - j)!}.
\end{align}
%
  
We denote our proposed prior as $HIW_c(\delta,W,G)$ and specify
\begin{align}
S_k \mid G \sim HIW_c(\delta_k,W_k,G) \quad k=0,\dots,T-1.
\end{align}
In the Supplement, we show that this prior specification is \emph{conjugate} to the TGM-based Whittle likelihood of Eq.~\eqref{eq:whittlegraph}. Also note that the graph, $G$, is shared across all frequencies.

\subsection{Marginal likelihood} \label{marglik}
Due to conjugacy of our proposed hyper complex inverse Wishart prior, the marginal likelihood of the time series $\BF{X}_{1:N}$ given a decomposable graph $G$, integrating out the spectral density matrices $S_{0:T-1}$, has a closed form which is derived in the Supplement and given by 
\begin{equation}\label{eq:marg}
p(\BF{X}_{1:N}|G) \approx \pi^{-N T p} \prod_{k = 0}^{T - 1}  \frac{h(W_k,\delta_k,G)}{h(W^*_k, \delta_k^*,G)}.
\end{equation}
Here, $\delta_k^* = \delta_k + N$, $W^*_k = W_k + P_k$, and
\begin{equation}
h(W,\delta,G) = \frac{\prod_{C \in \mathcal{C}} B(W_C,\delta)}{\prod_{S \in \mathcal{S}} B(W_S, \delta)}.
\end{equation}
From the definition of $\delta_k^*$, we see that $N$, the number of time series, acts as the effective number of observations in this case.  For the i.i.d. GGM, $N$ represents the number of independent vector-valued observations; in our TGM, $N$ plays the same role, but represents the number of independent \emph{time series} observations.  Likewise, as in standard inverse Wishart based modeling of covariances for i.i.d. Gaussian data, based on a set of $N$ i.i.d. complex normal observations of Fourier coefficients $d_{nk}$ with covariance $S_k$ (see Eq.~\eqref{eq:whittle}), we update the prior scale matrix $W_k$ with the outer product $P_k = \sum_{n=1}^N d_{nk}d_{nk}^*$, which is the aggregate \emph{periodogram} (see Eq.~\eqref{eq:periodogram}).

Having an analytic marginal likelihood of the time series given a PGM allows us to perform inference directly over graphs, sidestepping any thorny issues with inference directly on the $T$ $p \times p$ spectral density matrices themselves.  This is a critical feature of the practicality of our approach.
 
\subsection{Fractional priors for model selection}
Marginal likelihoods used for model comparison \cite{Kass:1995} are notoriously sensitive to the choice of prior parameters, or \emph{hyperparameters}. In our case, the marginal likelihood in Eq.~\eqref{eq:marg} depends strongly on the hyper complex inverse Wishart scale matrix, $W_k$. Since the scale and shape of the spectral density matrices are not known a priori, and vary dramatically across frequencies, we employ \emph{fractional priors} \cite{Ohagan:1995} over each $S_k$. Fractional priors effectively hold out some fraction of the data, and utilize that fraction to determine an adequate hyperparameter setting for each model. The rest of the data are then used for model comparison. Fractional priors have been deployed for graphical model selection in i.i.d. graphs and have a number of desirable properties such as information consistency and demonstrated robustness \cite{Scott:2008}. In our case, under a fractional prior with parameter $g \in (0,1)$, the fractional marginal likelihood is
\begin{equation} \label{eq:fracmarg}
p(\BF{X}_{1:N}|g, G) = \pi^{-N T p} \prod_{k = 0}^{T - 1} \frac{h(g P_k,g N,G)}{h(P_k, N,G)}.
\end{equation}
Here, we see that $g$ controls the fraction of data used for prior formulation versus model comparison.  Importantly, we now have just a single, scalar, and interpretable parameter $g$ to tune.  Default settings are suggested in \cite{Ohagan:1995,Scott:2008}.

\subsection{Graph prior}
There are two common approaches in the literature to specifying a prior distribution on graphs. The first approach places a uniform distribution on the space of all possible graphs \cite{Giudici:1999, Dellaportas:2003, Roverato:2002}. As noted in \cite{Armstrong:2009}, this prior puts high weight on graphs with a medium number of edges and significantly less weight on graphs with small or many edges. In response to this problem, it has been proposed to place a prior directly on the size of the graph and then consider a conditionally uniform prior on all graphs of the same size \cite{Armstrong:2009, Dobra:2004, jones:2005}. We follow this later approach and place a binomial distribution on the number of edges, $k$:
\begin{equation}
p(G) \propto r^k (1 - r)^{m - k}, 
\end{equation}
where $r$ is the prior probability that each of $m = \frac{p(p-1)}{2}$ possible undirected edges $(i,j) \in V \times V$ is included. Since $r$ is unknown, we further place a $\text{Beta}(a,b)$ prior over $r$. Integrating out $r$ gives the marginal prior over graphs
\begin{align} \label{eq:multprior}
p(G) \propto \frac{\beta(a + k, b + m - k)}{\beta(a,b)}
\end{align}
where $\beta(.,.)$ is the beta function. As explored in \cite{Scott:2008}, this is a multiplicity correcting prior \cite{Scott:2006} over graphs with the desirable property of diminishing false positive edge discoveries as extra unconnected nodes are added to the graph.

\section{METHODS FOR SINGLE TIME SERIES} \label{Sec:smoothing}
In some applications of interest one observes only a single multivariate time series, $N = 1$, from which the graph must be inferred. Two challenges arise in this setting: (1) the effective number of observations informing Eq.~\eqref{eq:marg} is just one and (2) the periodogram used in computing $W_k^*$ is noisy regardless of the length of the series, $T$.  The periodogram is a notoriously poor estimator of the spectral density, and when the spectral density itself is of primary interest, a common frequentist method is to smooth the periodogram to obtain a consistent spectral density estimator \cite{Jung:2014, Bach:2004, Dahlhaus:2000}. One could imagine using the smoothed periodogram as a plug-in estimator in Eq.~\eqref{eq:marg}, scaled by the effective degrees of freedom (see the Supplement for more details on this plug in estimator for our formulation).  An alternative variance-reduction technique is the Bartlett method \cite{Bartlett:1948}, that divides the length $T$ series into $M$ shorter series of length $\frac{T}{M}$ and averages the resulting $M$ periodograms, but at the cost of reduced resolution (i.e., number of considered frequencies).  This approach mimics the implicit smoothing that occurs when we compute the periodogram based on $N$ truly independent series each of length $T$, as in Eq.~\eqref{eq:periodogram}.

In contrast to a plug-in estimator, a natural Bayesian approach enforces smoothing across frequencies via a prior distribution over the set of spectral densities \cite{Rosen:2007}. Previous approaches have coupled elements of a Cholesky decomposition of each spectral density matrix across frequencies, however this approach is unsuitable to our case since 1) it does not enforce sparsity in the inverse spectral density and 2) a prior of this form will remove the simple marginal likelihood structure in Eq.~\eqref{eq:marg} that we harness for efficient inference. Motivated by our aims to both share information across frequencies and maintain the form of the marginal, we utilize a piecewise constant prior over spectral densities given a graph, $G$. We partition the interval $[0, 2 \pi]$ into $M$ intervals $w_1 = \left[0, \frac{2 \pi}{M}\right), \ldots, w_j = \left[\frac{2 \pi (j - 1)}{M}, \frac{2 \pi j}{M}\right),\ldots, w_{M}=\left[\frac{2 \pi (M - 1)}{M}, 2 \pi \right]$ and then draw a separate positive definite Hermitian matrix from a $HIW_c$ distribution for each interval:
\begin{align}
\tilde{S}_j &\sim  HIW_{c}(\delta,W_j, G) \quad j=1,\dots,M.
\end{align}
Our resulting spectral density is simply
\begin{align}
S(\lambda) &= \sum_{j = 1}^{M} {\BF 1}_{\lambda \in w_{j}} \tilde{S}_j \quad \forall \lambda \in [0, 2 \pi].
\end{align} 
Under this prior, the marginal likelihood for the single ($N = 1$) time series becomes
\begin{equation}
p(\BF{X}|G) \approx \pi^{-M p} \prod_{j = 1}^{M} \frac{h(W_j,\delta_j,G)}{h(W^*_j, \delta^*_j,G)}
\end{equation}
where $\delta^*_j = \delta_j + \sum_{k = 0}^{T-1} {\BF 1}_{\lambda_k \in w_{j}} $ and $W^*_j = W_j + \sum_{k = 0}^{T-1} {\BF 1}_{\lambda_k \in w_{j}} P_k$. By setting $M = \lfloor \sqrt{T} \rfloor$, we obtain an asymptotically approximate nonparametric prior distribution over continuous spectral density matrices: for $T$ large enough the prior puts positive support on spectral density matrices arbitrarily close to any continuous spectral density over $[0, 2 \pi]$. Furthermore, under this setting as $T \to \infty$, the number of Fourier frequencies, and thus number of samples $\sum_{k = 0}^{T-1} {\BF 1}_{\lambda_k \in w_{j}}$, within each interval grows as $\sqrt T$. 

\section{INFERENCE}\label{sec:inference}
Bayesian structural learning algorithms for decomposable graphs come in two flavors: MCMC samplers and stochastic search procedures \cite{Scott:2008, Moghaddam:2009}.  By placing decomposable graphical inference for time series in the same framework as for the i.i.d. case via our analytic $p(\BF{X}_{1:N}\mid G)$, we can easily modify both types of existing methods for the time series case. 

Classical MCMC samplers for decomposable graphs sample from the posterior over graphs via Metropolis-Hastings (MH) by proposing single edge addition and deletion moves that keep the graph decomposable \cite{Giudici:1999, Armstrong:2009}. While it is possible to obtain any decomposable graph from any other decomposable graph via a sequence of edge additions and deletions, the path may be hard to reach leading to prohibitive converge times for even a moderate number of vertices $p$. More recent graph samplers add more global moves by either randomly generating new decomposable graphs \cite{Zhu:2014} or by generating from a Markov chain over a junction tree representation of the graph \cite{Green:2013}. To compute the MH acceptance ratio, these samplers rely on computing ratios of present and proposed marginal likelihoods. For simple edge additions and deletions, this ratio simplifies into a function of only the cliques and separators that change between moves.  For our case, the ratio expands into a product over frequencies of the same affected cliques and separators, allowing simple modifications to the existing implementations of these samplers to handle TGMs.

All current MCMC samplers struggle to scale to even moderate numbers of nodes. In contexts where point estimates suffice, we can instead consider stochastic search procedures.  We utilize a modification of the efficient feature-inclusion stochastic search (FINCS) \cite{Scott:2008} for inference in our TGMs.  
FINCS interleaves three moves: 1) single edge addition and deletion moves for local changes to the graph, 2) global sampling moves where edges are added independently to an empty graph and the final graph is triangulated to maintain decomposability, and 3) resampling at step $t$ a full graph from a list of past visited models, $\{G_1, G_2, \ldots, G_{t-1}\}$, in proportion to their posterior probabilities. In steps 1) and 2), to enforce exploration of high probability regions, edge additions that tend to continually improve the model probability are preferentially selected in proportion to a current heuristic estimate of the posterior edge probability
\begin{equation}
\hat{q}_{ij}(t) = \frac{\sum_{k = 1}^{t} 1_{\{i,j\} \in E_t} p(X_{1:N}|G_{t}) p(G_{t})}{\sum_{k = 1}^{t} p(X_{1:N}|G_{t})},
\end{equation}
where $E_t$ is the current edge set. 
Edge deletions are performed proportional to $\hat{q}_{ij}^{-1}(t)$. As in MCMC samplers \cite{Giudici:1999, Armstrong:2009}, the junction tree representation of the graph can be efficiently updated after each local move since the two graphs only differ by a single clique and its corresponding separators, allowing a quick computation of the marginal likelihood of a proposed graph in Eq.~\eqref{eq:marg}.  Importantly, the FINCS algorithm depends on the data only through the marginal likelihoods of the cliques $C$---used to compute the full graph marginal likelihood---which in our TGM case is a product over $T$ frequencies:
\begin{equation}
\prod_{k = 0}^{T - 1} \frac{B(W_{k,C}, \delta)}{B(W^*_{k,C}, \delta^*)}.
\end{equation}  
That is, our implementation simply modifies the original FINCS definition of the clique marginal likelihood.

\section{SIMULATIONS}\label{sec:sim}
To test our TGM methods, we consider simulated setups for both $N>1$ and $N=1$ time series each generated from an order-1 vector autoregressive process, denoted VAR(1), for $p=20$ dimensions. Specifically, we simulated data from the model
\begin{equation}
x(t) = A x(t - 1) + \epsilon(t),
\end{equation}
where $x(t) \in \mathbb{R}^p$, $A \in \mathbb{R}^{p \times p}$, and
$\epsilon(t) \sim N(0,I_{p \times p})$. The inverse spectral density of a
VAR(1) process is given by \cite{Songsri:2011}
\begin{equation} \label{eq:specinv}
S(\lambda)^{-1} = I + A^T A + e^{-i \lambda} A + e^{i \lambda} A^T.
\end{equation}
Random sparse TGMs were generated by first restricting $A$ to be upper
triangular. Following the simulated setup in \cite{Songsri:2011}, we set the diagonal elements to a constant $A_{ii} = .5$ and sample the upper diagonal elements as $a_{ij} \sim .5 \delta_{ij}$, where $\delta_{ij} \sim \text{Binomial}(\rho)$ with $\rho = .2$ for all simulations.  The graph $G$ was then determined by identifying the zeros in $S(\lambda)^{-1}$ using Eq.~\eqref{eq:specinv}. Proposed $A$ matrices were accepted when both the absolute value of all eigenvalues of $A$ were less than one, making the series stationary, and the graph $G$ determined by $A$ was decomposable.

We note that since our formulation reduces to a standard structure learning problem, our emphasis is less on assessing performance with respect to $p$, which should follow from whichever structure learning algorithm is selected; instead, our focus is on $N$ and $T$, which are specific to the time series and spectral analysis.  For example, in the FINCS algorithm \cite{Scott:2008}, it is quoted that the method can handle graphs with up to roughly $p=100$ nodes.

\begin{figure*}[t]
        \centering
		\begin{tabular}{cccc}
					\includegraphics[width=0.25\textwidth]{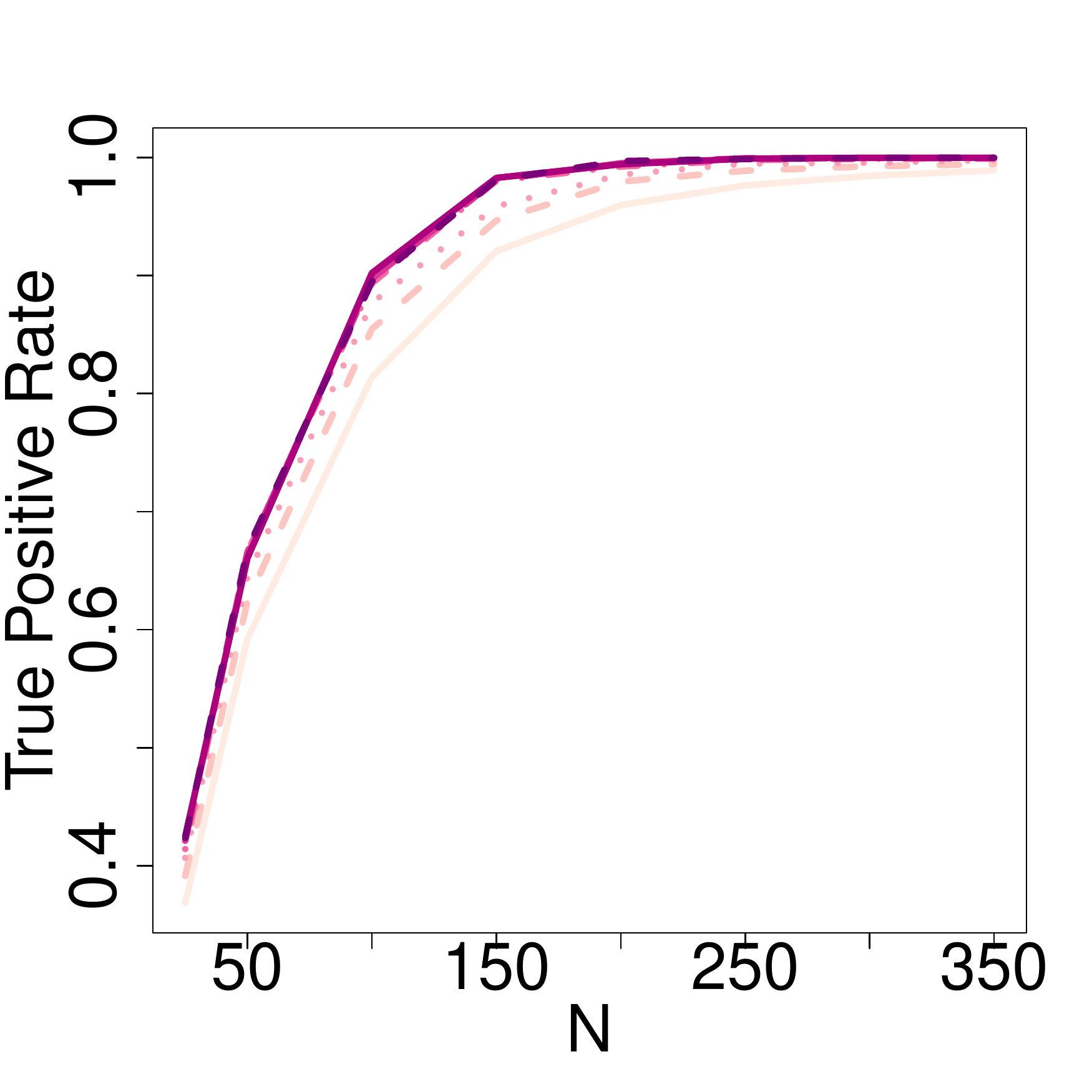}
	        &
	                \includegraphics[width=0.25\textwidth]{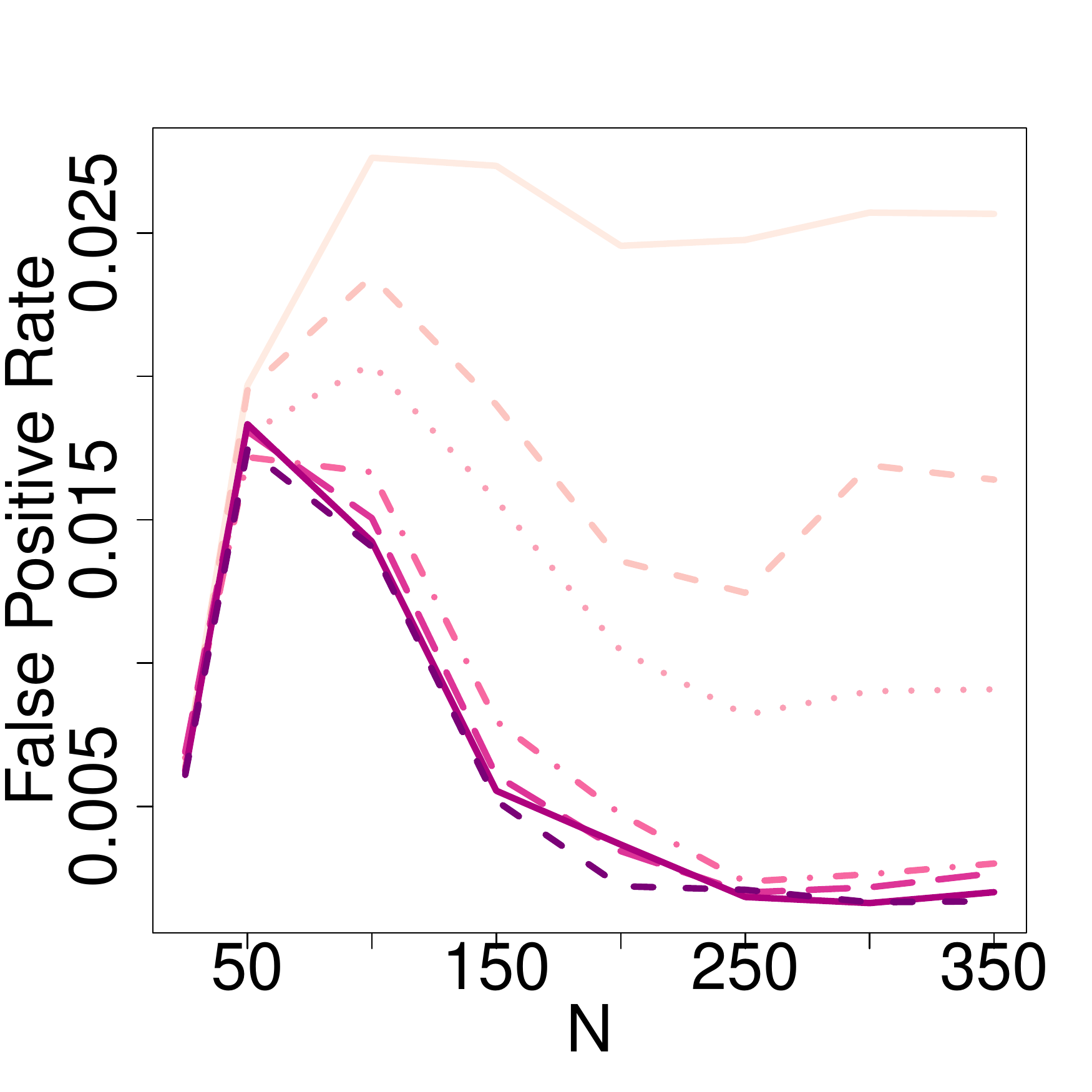}
	       	&
	                \includegraphics[width=0.25\textwidth]{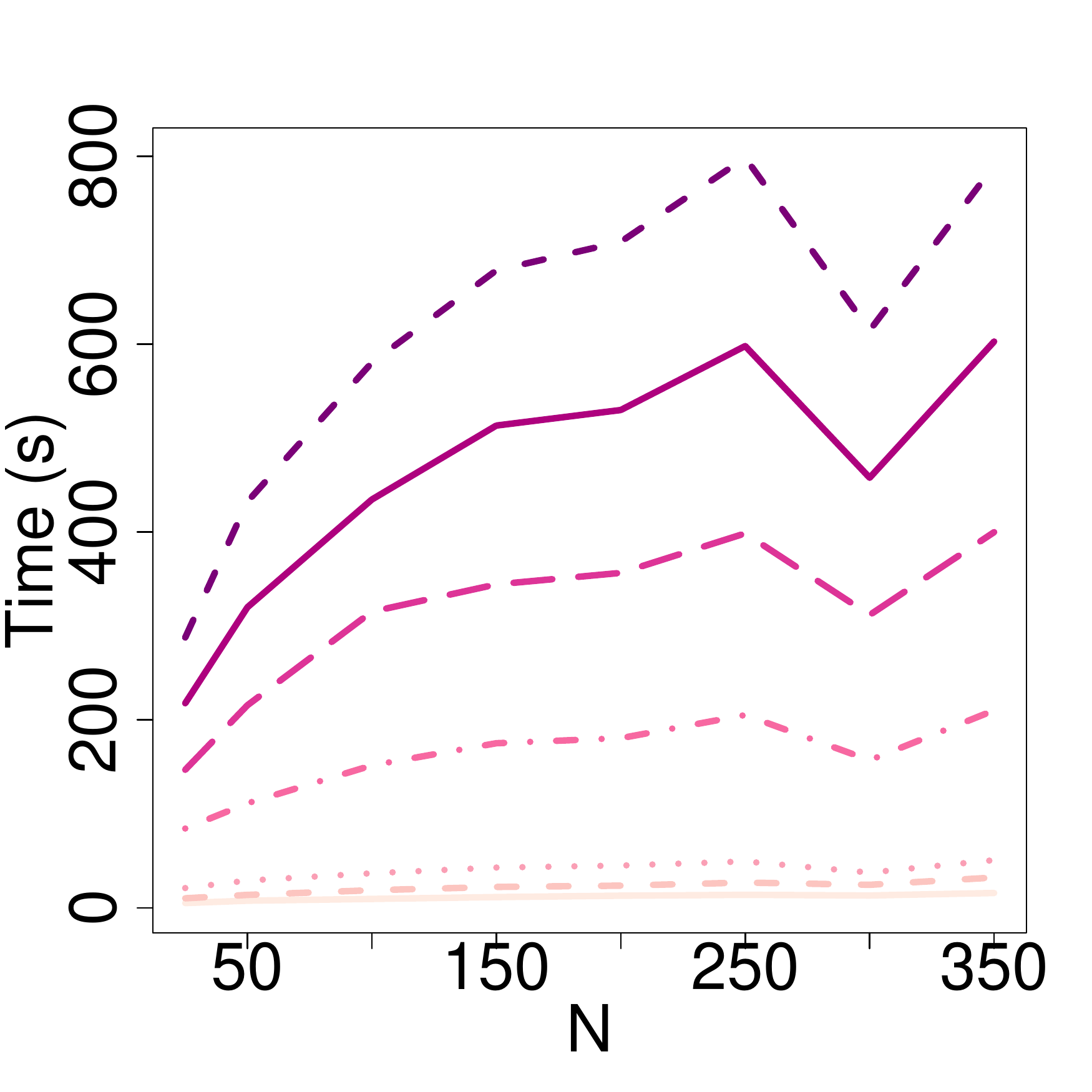} 
	        &     \hspace{-.2in}   \includegraphics[width=.1\textwidth]{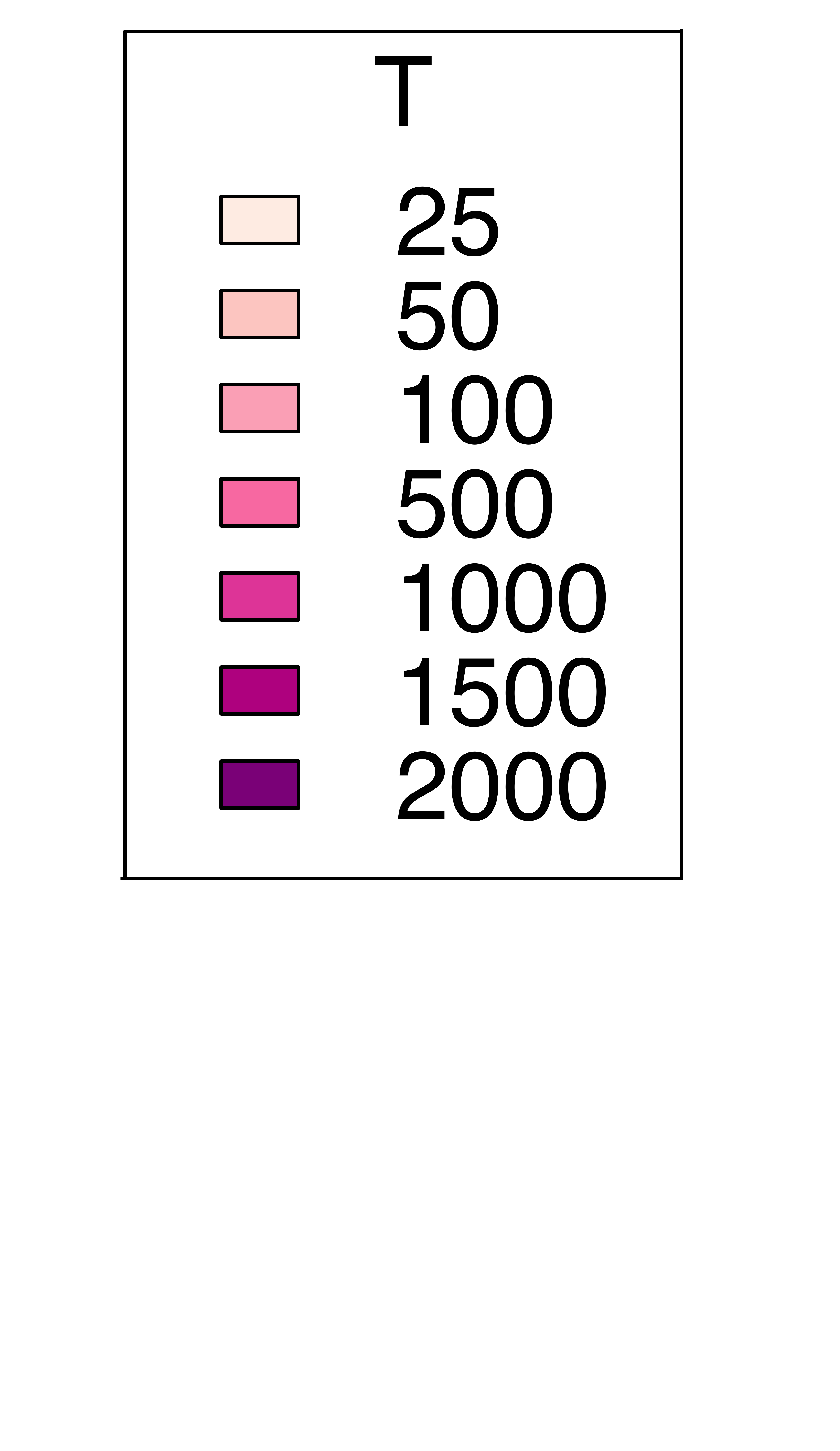} \hspace{-.2in}
	                	\vspace{-0.2in}\\
                \includegraphics[width=0.25\textwidth]{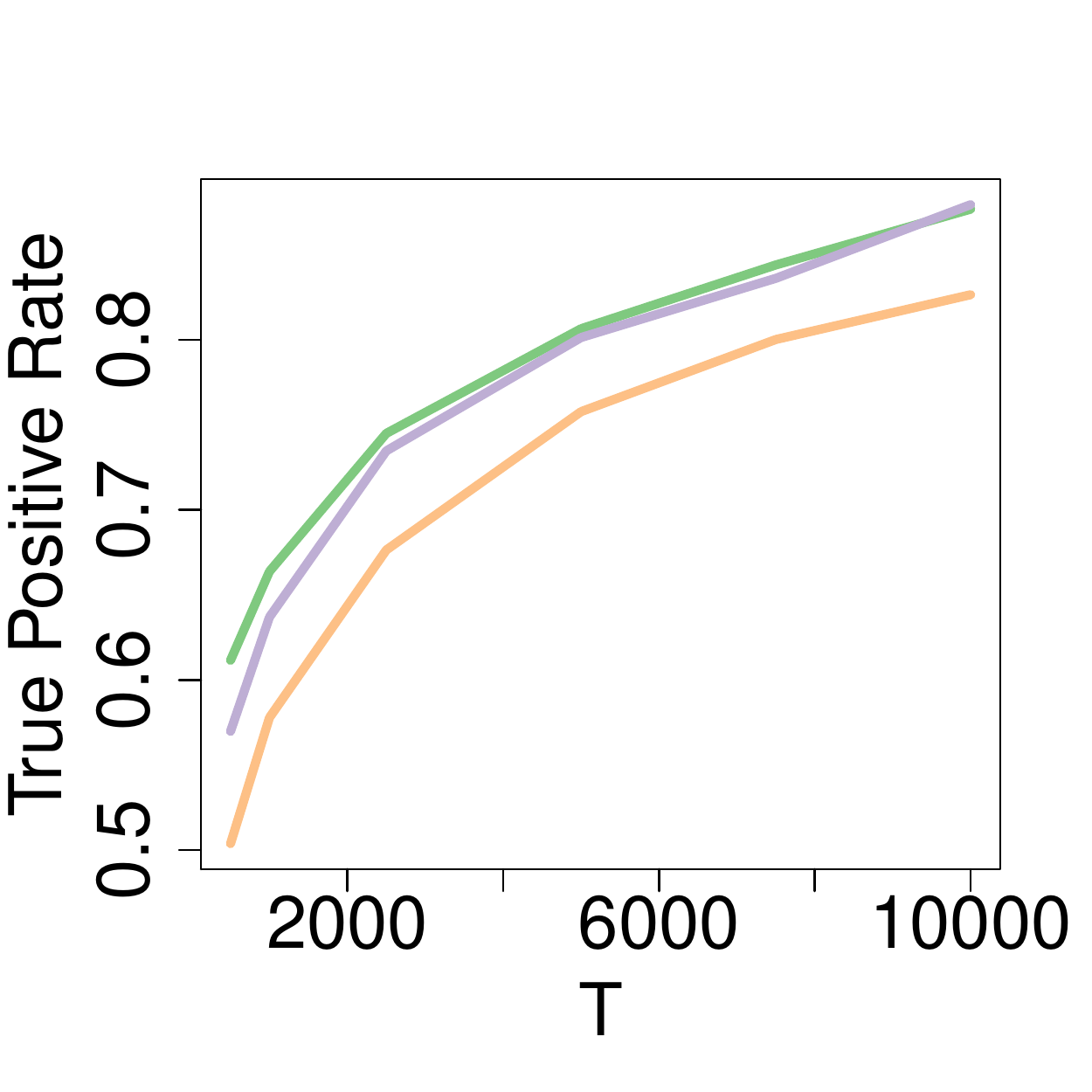}
        &
                \includegraphics[width=0.25\textwidth]{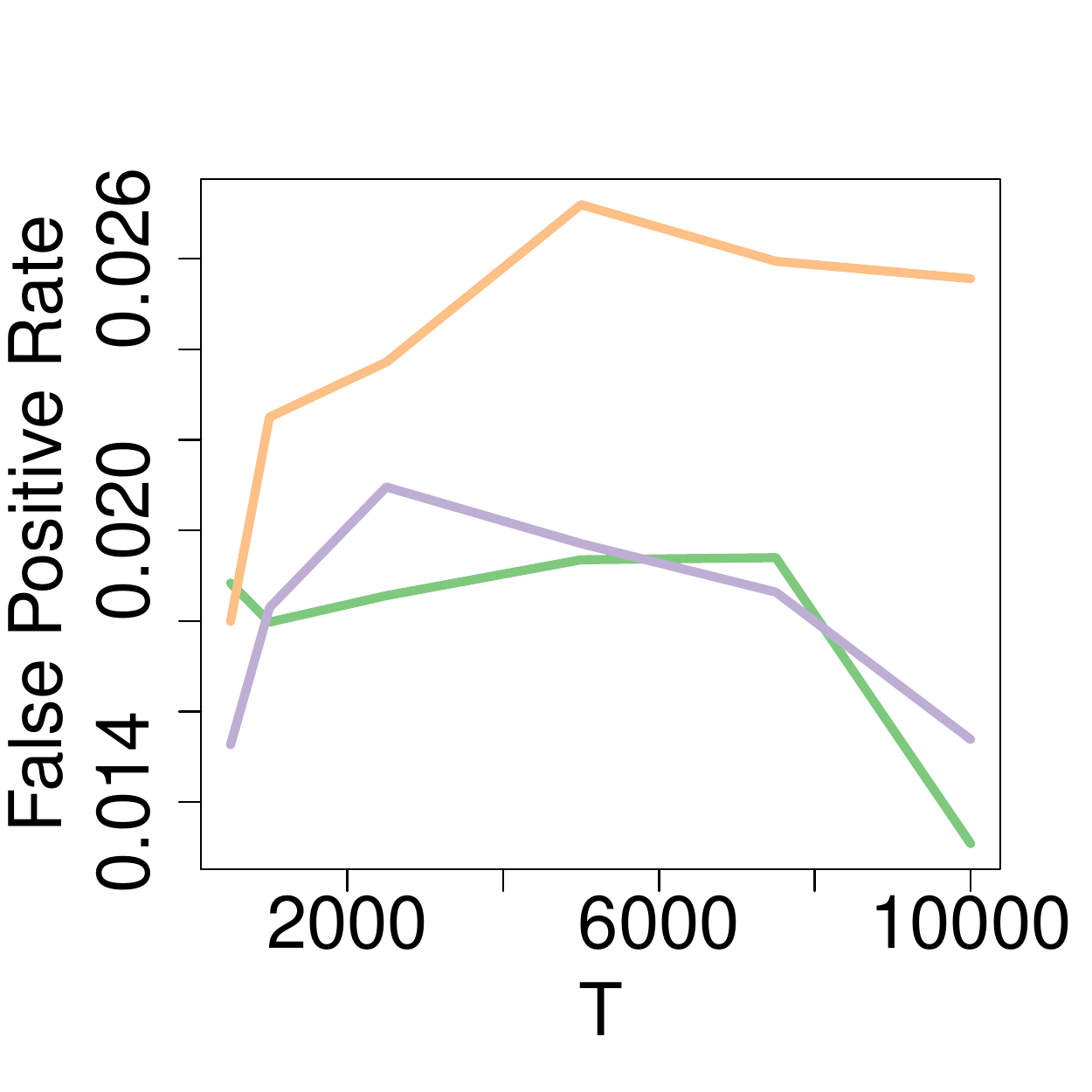}
        &
                \includegraphics[width=0.25\textwidth]{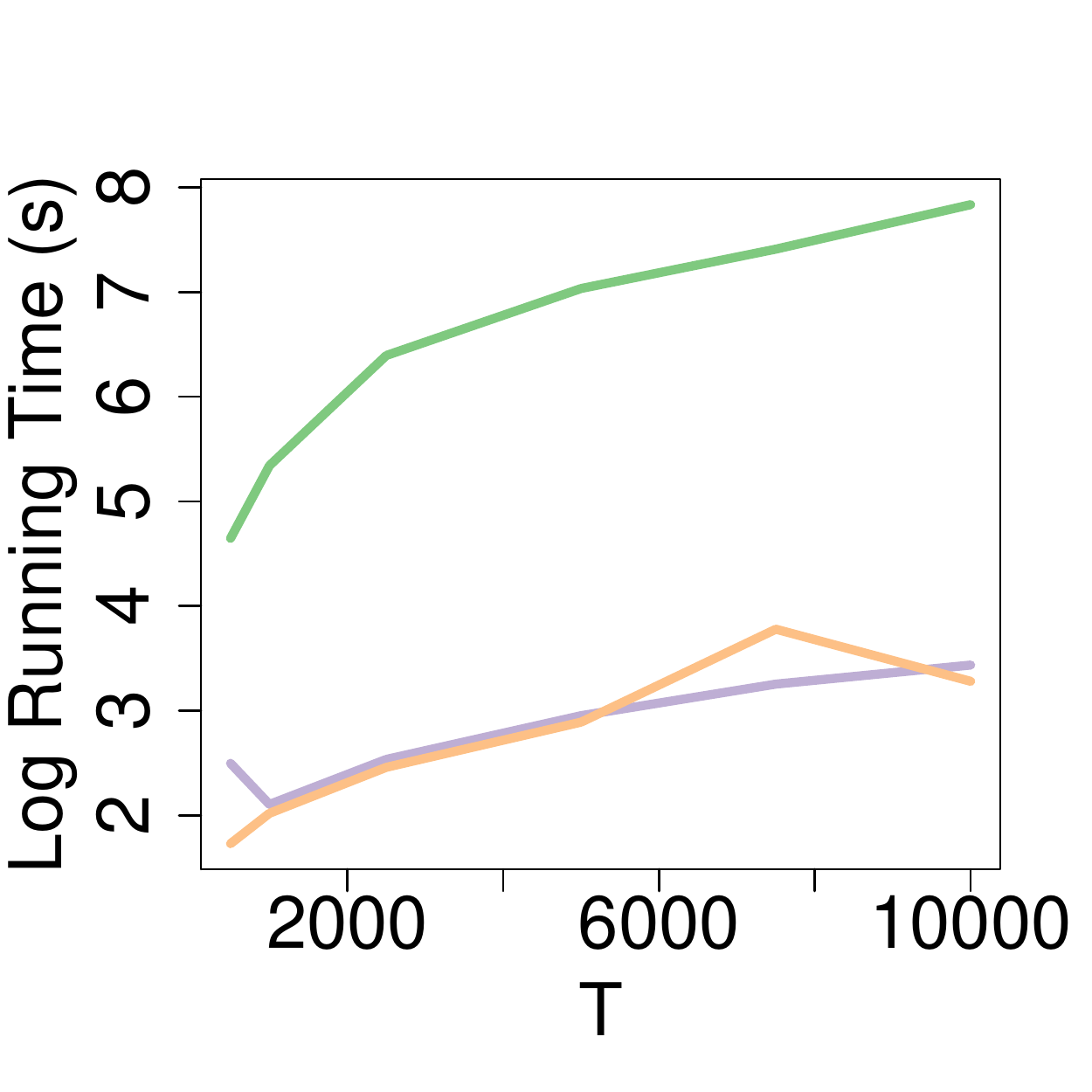}
        &       
        		\hspace{-.2in}	\includegraphics[width=.1 \textwidth]{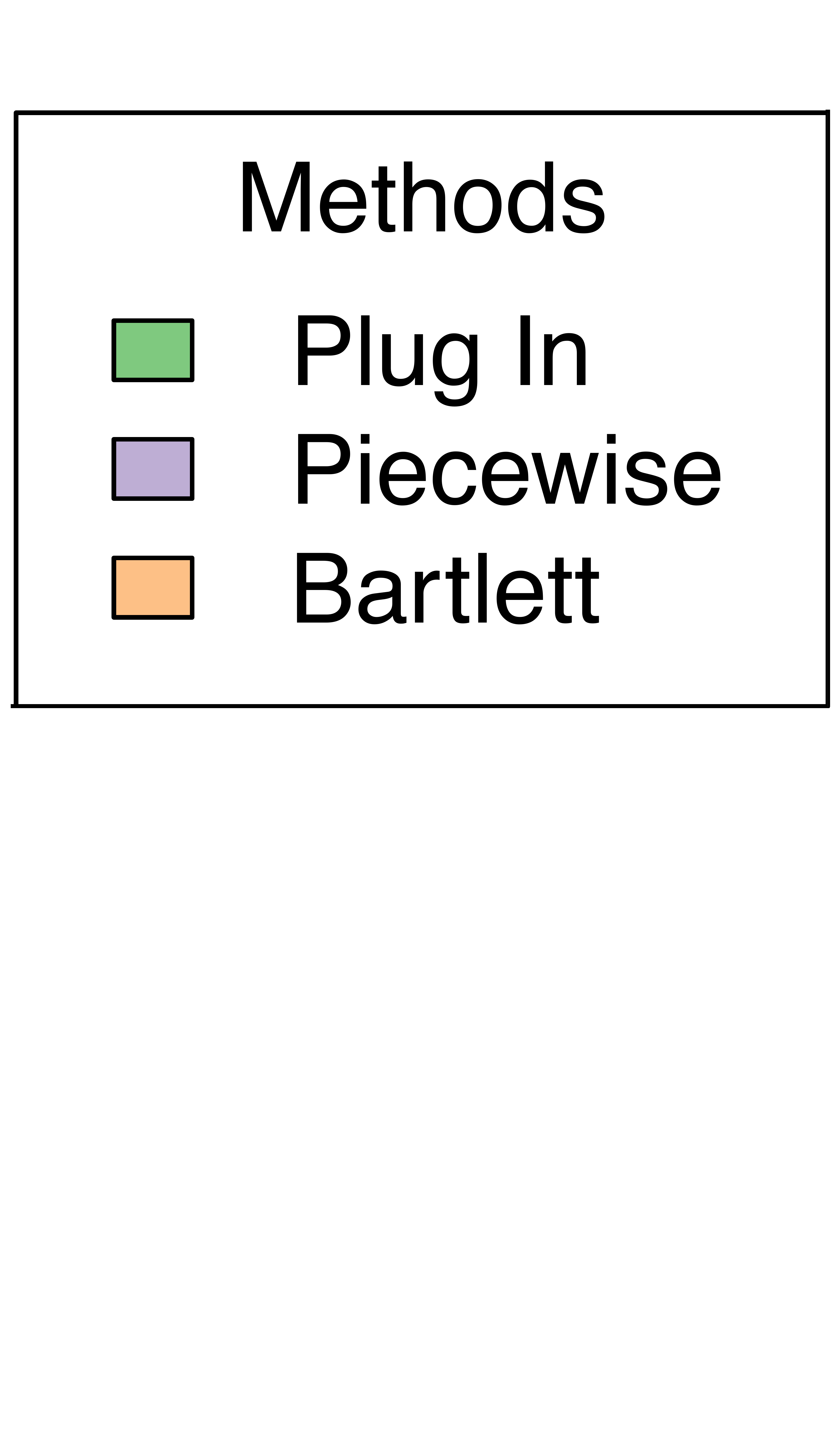}    \hspace{-.2in}
		\end{tabular}
		\vspace{-0.1in}
        \caption{{\bf Top:} As a function of the number of time series $N$, and plotted for various values of their length $T$, (\emph{left}) mean true positive rate,  (\emph{middle}) median false positive rate, and (\emph{right}) mean running time computed across the 200 replicates.  Standard error bars are small relative to the scale of the plots and are omitted for clarity.  {\bf Bottom:} Same plots as a function of $T$ for a single time series ($N=1$), and plotted for various periodogram smoothing techniques.}
        	\label{fig:sim}
\end{figure*}
\subsection{Multiple time series}
To analyze how our TGM structure learning performance varies with the number of time series replicates, $N$, we simulated data for $N \in \{20, 50, 100, 150, 200, 250, 300,350\}$ and $T \in \{25,50,100 500, 1000, 1500, 2000\}$. This process was repeated $200$ times for each combination of $N$ and $T$. Each time series is first decomposed into its discrete Fourier components. We then ran 10,000 iterations of the FINCS algorithm using the fractional marginal likelihood in Eq.~\eqref{eq:fracmarg} with $g = \frac{4}{N}$, a default setting \cite{Ohagan:1995,Scott:2008}. Our graph prior followed the multiplicity correcting form in Eq.~\eqref{eq:multprior} with $a = b = 1$. The graph visited with highest posterior probability was then selected and true and false positive rates were computed. Results are displayed in Fig.~\ref{fig:sim}. Across $T$, the true positive rate increases quickly with the number of series, $N$, achieving an almost perfect true positive rate by about $N = 150$. We also see that the rate of increase in the true positive rate increases with the length of the series $T$, which relates to the number of considered Fourier frequencies. It is interesting to note that for all $T$ under consideration, the false positive rate tends to start very low $(\approx.005)$ for $N = 20$ replicates then spike at $N \in \{50, 100\}$ before declining again. This occurs due to the fact that at low $N$, very few edges are introduced at all, perhaps due to an Occam's razor type effect of marginal likelihoods penalizing model complexity. As $N$ starts to increase, more edges are introduced, both correct and incorrect, and as $N$ further increases, the false edges are pruned and true edges are retained, leading to a decline in the false positive rate. Note that the false positive spike tends to be more pronounced for time series of smaller length, $T \in \{25,50\}$. One would expect to see significant improvements, especially for small $N$, by leveraging the piecewise constant prior of Sec.~\ref{Sec:smoothing} and explored in Sec.~\ref{sec:singleTS} where we show that we are able to learn graphs from just $N=1$ time series.  However, we chose not to include this prior in this analysis so as not to confound its effect with our performance.  Here, the noisy periodogram is smoothed implicitly by averaging over $N$.

Finally, in Fig.~\ref{fig:sim} we see that runtime increases as a function of $T$ due to the dependence on $T$ in the marginal likelihood computation of Eq.~\eqref{eq:marg}, though significant cost reductions can be achieved through parallelizations leveraging the product form.  

\subsection{Single time series: comparison of methods}\label{sec:singleTS}
\begin{figure*}[t]
        \centering
		\begin{tabular}{cccc}
                \includegraphics[width=0.2\textwidth]{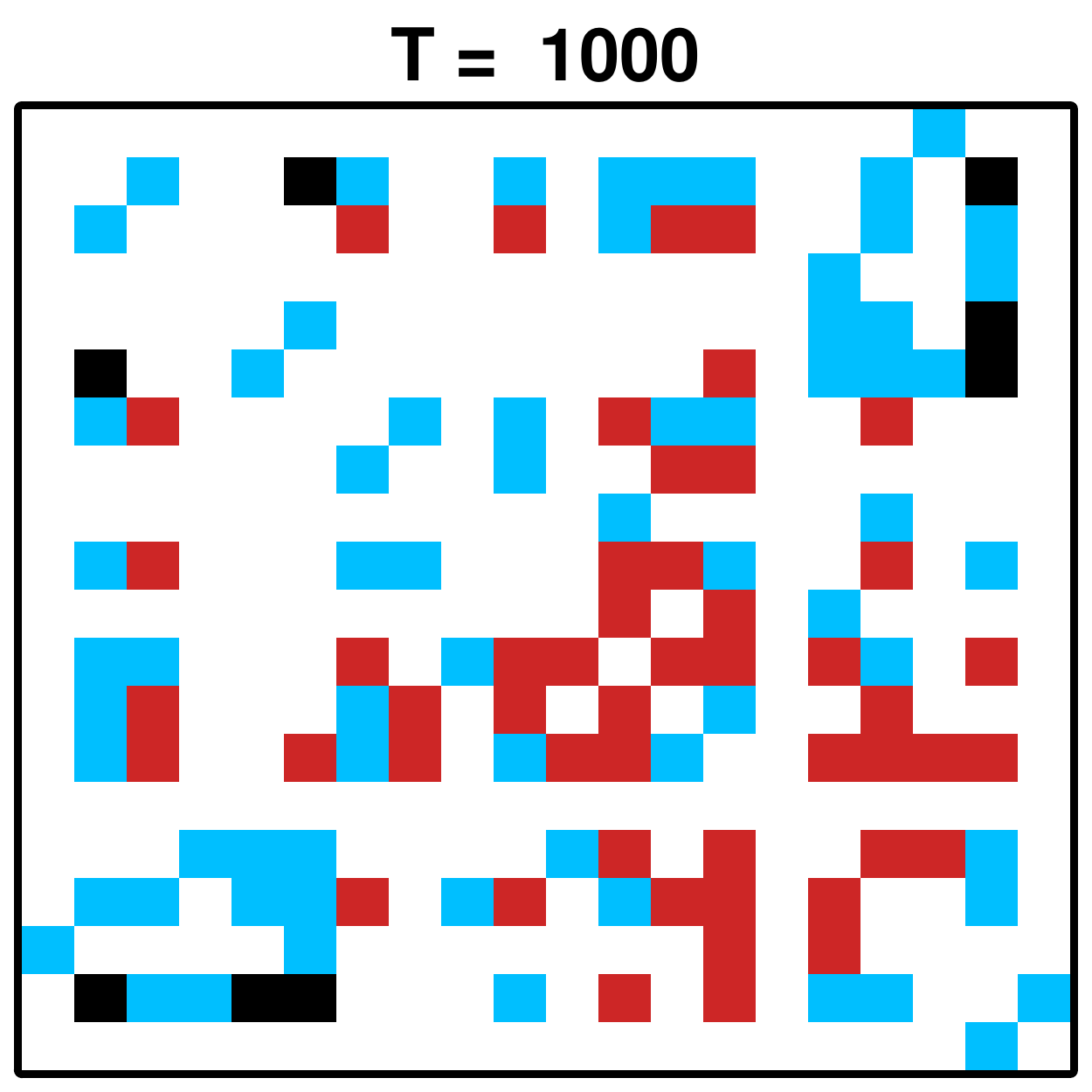}
        &
                \includegraphics[width=0.2\textwidth]{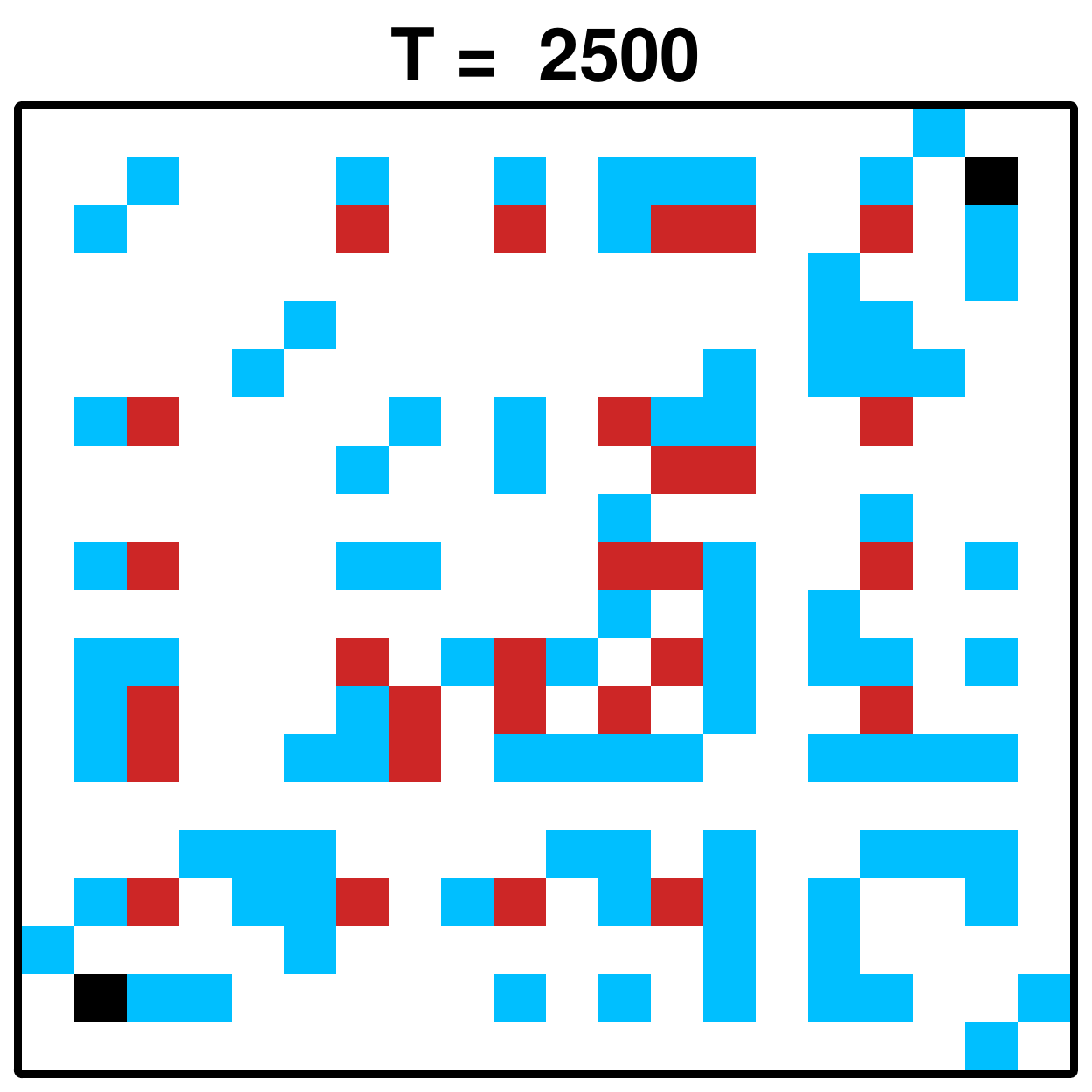}
        &
                \includegraphics[width=0.2\textwidth]{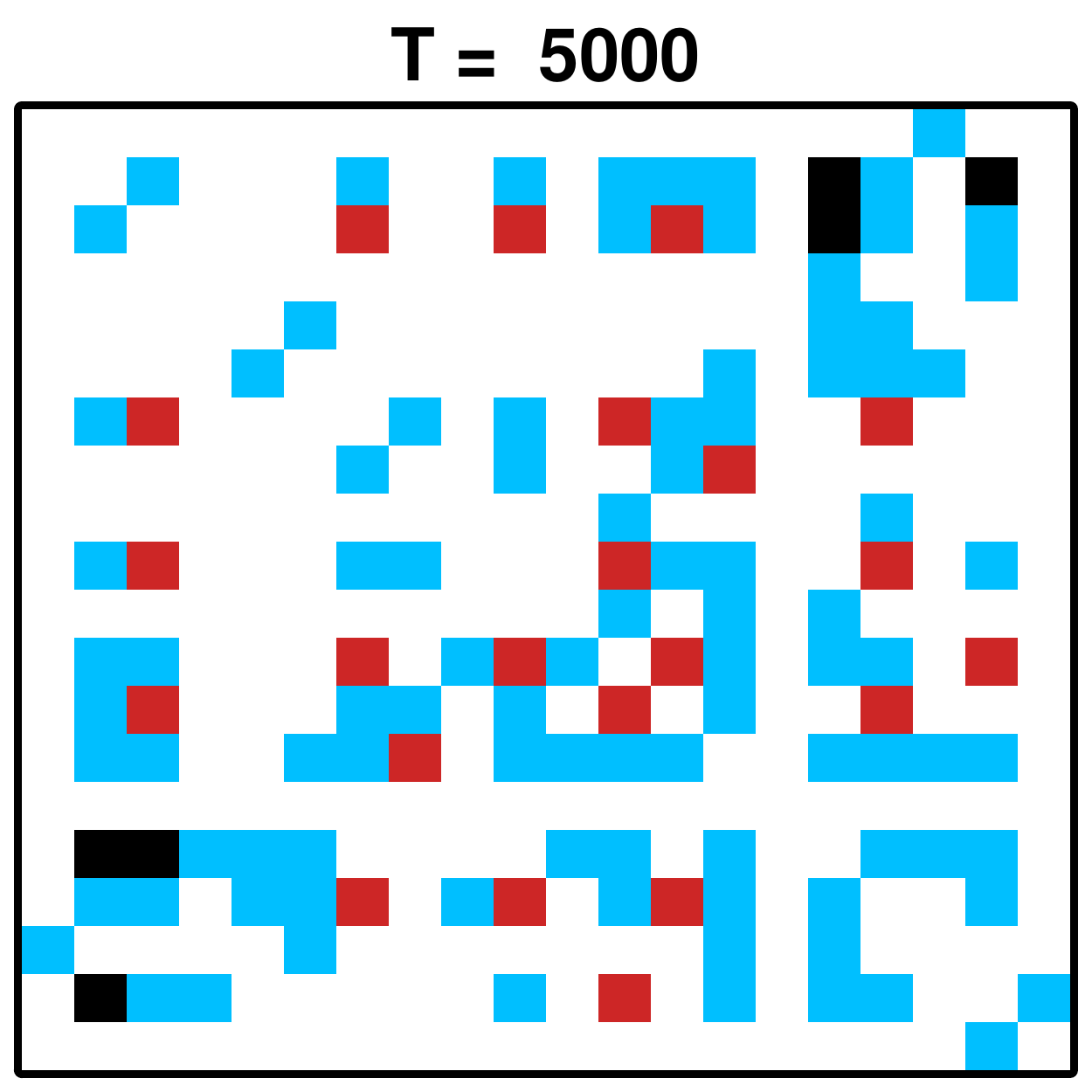}
        &
		\includegraphics[width=0.2\textwidth]{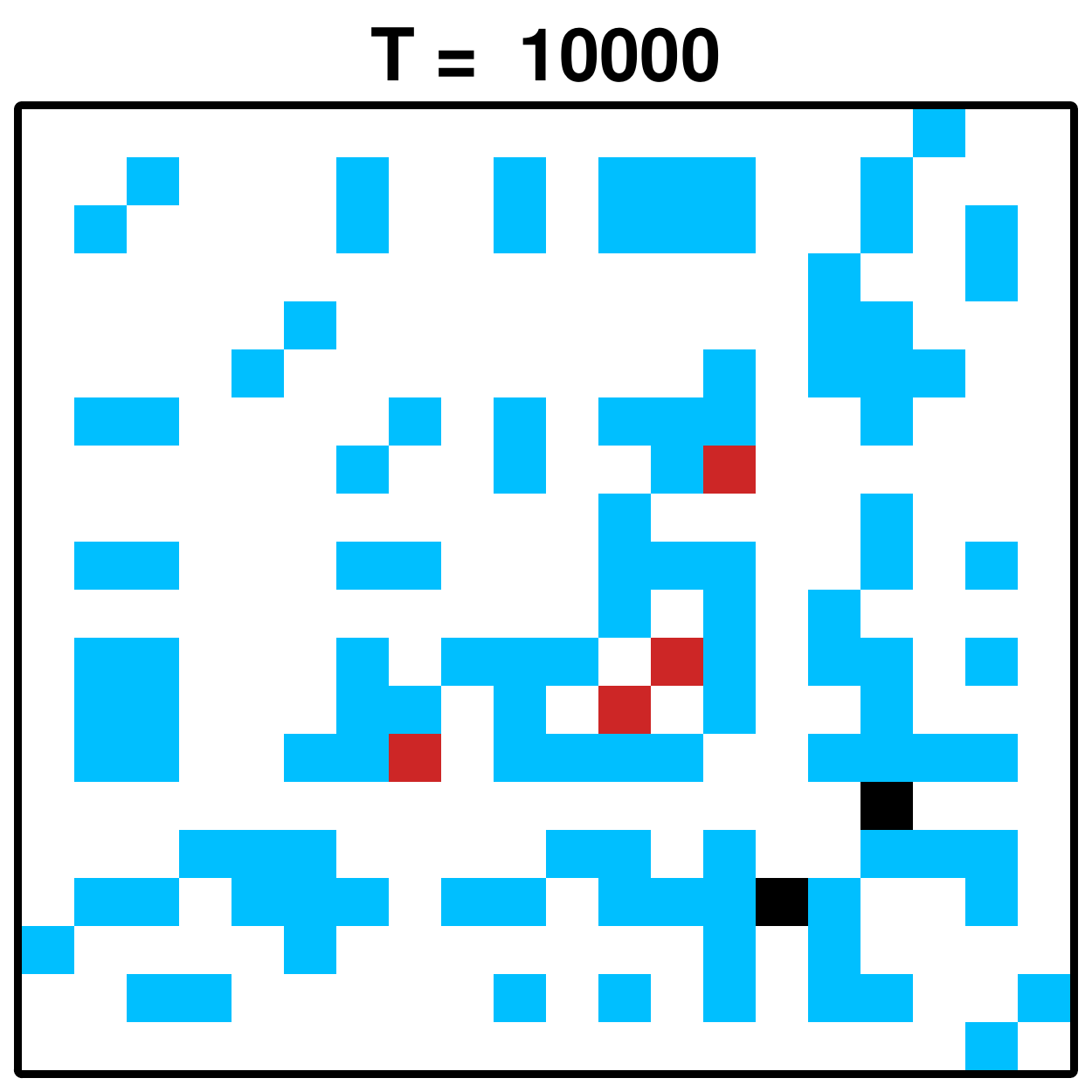}
		\end{tabular}

		\caption{ Example evolution of error types for the piecewise prior method as a function of series length, $T \in \{1000,2500,5000,10000\}$ and $N = 1$, for a selected graph. \textcolor{blue}{Blue}, \textcolor{red}{red}, {\bf black}, and white entries indicate true positives, false negatives, false positives, and true negatives, respectively. The graph was selected by choosing the graph out of 200 replications with median true positive rate at $T = 2500$.}\label{fig:mediangraph}
		
\end{figure*}

To assess the performance of our single-time-series methods outlined in Sec.~\ref{Sec:smoothing}, we simulated a time series with $T \in \{500, 1000, 2500, 5000, 7500, 10000\}$. For the piecewise constant prior method, we use $M=\lfloor \sqrt{T} \rfloor$ pieces. We compare against the Bartlett time-series-splitting approach with the number of splits set to $\lfloor \sqrt{T} \rfloor$. We also examine a smoothed plug-in estimator of the spectral density using a Daniell smoother outlined in the Supplement with $m = \lfloor \frac{\sqrt{T}}{2} \rfloor$ for a total window size of $2 \lfloor \frac{\sqrt{T}}{2} \rfloor + 1 \approx \lfloor \sqrt{T} \rfloor$. For each method, the FINCS algorithm was run for 10,000 iterations and the highest scoring graph was selected and used to compute true and false positive rates. This process was repeated 200 times with results displayed in Fig.~\ref{fig:sim} with a replicate representative of our median performance shown in Fig.~\ref{fig:mediangraph}. The true positive rate increases for all three methods as a function of $T$, achieving a final value of about $.9$ for both the plug-in and piecewise constant prior methods and $.79$ for the Bartlett method at $T = 10000$. All methods maintain a low false positive rate around $.02$. Overall, the Bartlett method performs uniformly worse in terms of both true and false positive performance, while the piecewise prior method performs on par with the plug-in method, but at a fraction of the computational cost. 
Further experimental simulations are given in the Supplement. 

\vspace{-.13in}
\section{GLOBAL STOCK INDICES}\label{sec:stock}
\begin{figure}[t]
    \centering
	\begin{tabular}{cc}
        \hspace{-0.2in}\includegraphics[width=0.23\textwidth]{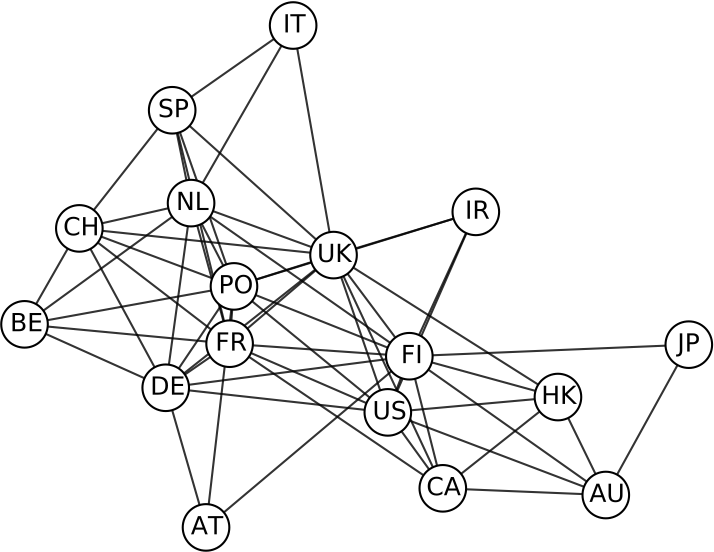}
        \includegraphics[width=0.23\textwidth]{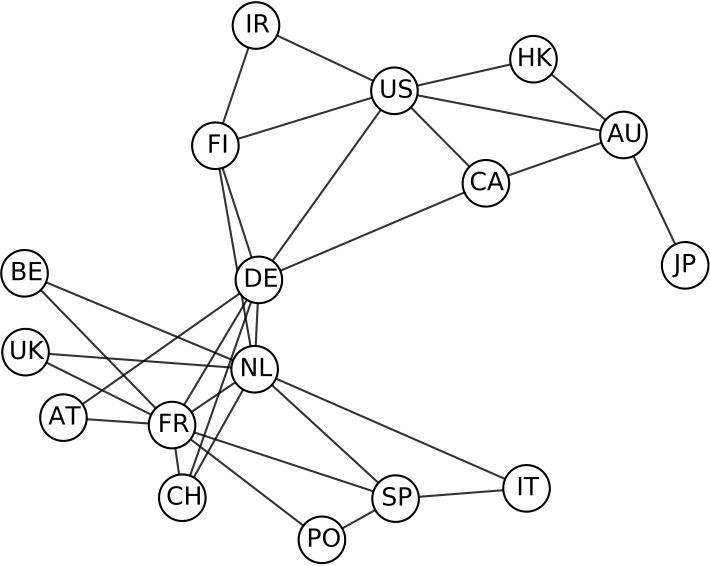}
    \end{tabular}
	\caption{Graphical models with the highest posterior probability for the stock index data. \textbf{Left:} Treating the log-returns as independent. \textbf{Right:} Using our TGM algorithm.  In both cases, we see regional connections, but our TGM algorithm results in a sparser and more interpretable graph.} \label{fig:stocks}
\end{figure}

\vspace{-.1in}
We explore the utility of our method in discovering conditional independencies
between countries inherent in the global financial system.  A similar
experiment was conducted in \cite{Songsri:2011} using a penalized-likelihood approach to learn TGMs, but restricted to finite-order VAR models with pre-specified order. (Recall that our method only assumes Gaussian stationarity, which includes the class of possibly infinite order VAR processes.)  Using \texttt{www.globalfinancialdata.com}, we acquired the
daily closing prices of 17 stock indices in US dollars for various countries around the world
(see the Supplement for the full list) from June 3, 1997 to June 30,
1999.  Missing prices were backfilled and only days where all exchanges traded
were considered which resulted in time series of length $542$.  Following
standard practice when analyzing stock prices, we converted
the closing prices, $p_t$, on day $t$ to log-returns according to \vspace{-.1in}
\begin{equation*}
    \label{eqn:logreturns}
    r_t = 100 \log(p_t / p_{t-1}).
\end{equation*}
We compare the graphical models inferred under two settings: (i) treating the log-returns
as independent (as in \cite{Scott:2008}) and (ii) using our methods to learn a TGM
treating the log-returns as a time series.  The best graphical models learned in each
scenario are depicted in Fig.~\ref{fig:stocks}.

For our TGM algorithm, we computed the periodogram for the 17-dimensional
time series, resulting in $542$ complex-valued matrices of dimension $17 \times
17$.  Since we only have one realization of the time series, we smoothed the
periodogram using the techniques and settings discussed in Sec.~\ref{sec:singleTS}.
We then ran the FINCS algorithm for 100,000 iterations.  
We compare the resulting highest-probability graph (see Fig.~\ref{fig:stocks}) to that learned treating the time series as independent based on the model in \cite{Scott:2008}, again using 100,000 iterations of the FINCS algorithm, but in its originally proposed form for non-temporal data.

In Figure~\ref{fig:stocks}, we see that in both cases we recover some geographical relationships between countries.  However, the independent model returns a significantly denser graph than that learned by our TGM approach.  Since the independent
model is not taking the temporal nature of the data into account, some edges
are likely spurious due to random correlations.  The TGM, on the other hand,
provides an interpretable and intuitive structure with strong geographic
connections. For example, there is a distinct United Kingdom / eurozone cluster
of Germany `DE', Finland `FI', Netherlands `NL', Belgium `BE', Switzerland
`CH', Austria `AT', Spain `ES', Italy `IT', Portugal `PT', and the United
Kingdom `UK'.  Another distinct cluster includes the United States `US', Canada
`CA', Hong Kong `HK' (whose currency is linked to the USD), and Australia `AU'
(whose currency is correlated with the US S\&P), with Japan `JP' hanging off
this cluster.  One perhaps strange missing link is between Ireland `IE' and the
UK, though the US and Ireland have a long history of economic connections
possibly explaining why Ireland is included in the separator between these two
distinct clusters.

In the Supplement, we include (i) a comparison of our learned graph with that of
Songsiri et. al.~\cite{Songsri:2011}, and (ii) further details on the stock
data itself.

\section{MAGNETOENCEPHALOGRAPHY DATA}\label{sec:meg}
\begin{figure*}[t!]
\vspace{-.2in}
    \centering
	\hspace{-0.25in}\begin{tabular}{c}
		\rotatebox{90}{\textbf{Intersection}} \vspace{0.5in}\\
		\rotatebox{90}{\textbf{Difference}} 
	\end{tabular}
	\begin{tabular}{ccccc}
        \textbf{High pitch (U)} & \textbf{Low pitch (D)} & \textbf{Left (L)} & \textbf{Right (R)} \\
        \includegraphics[width=0.17\textwidth]{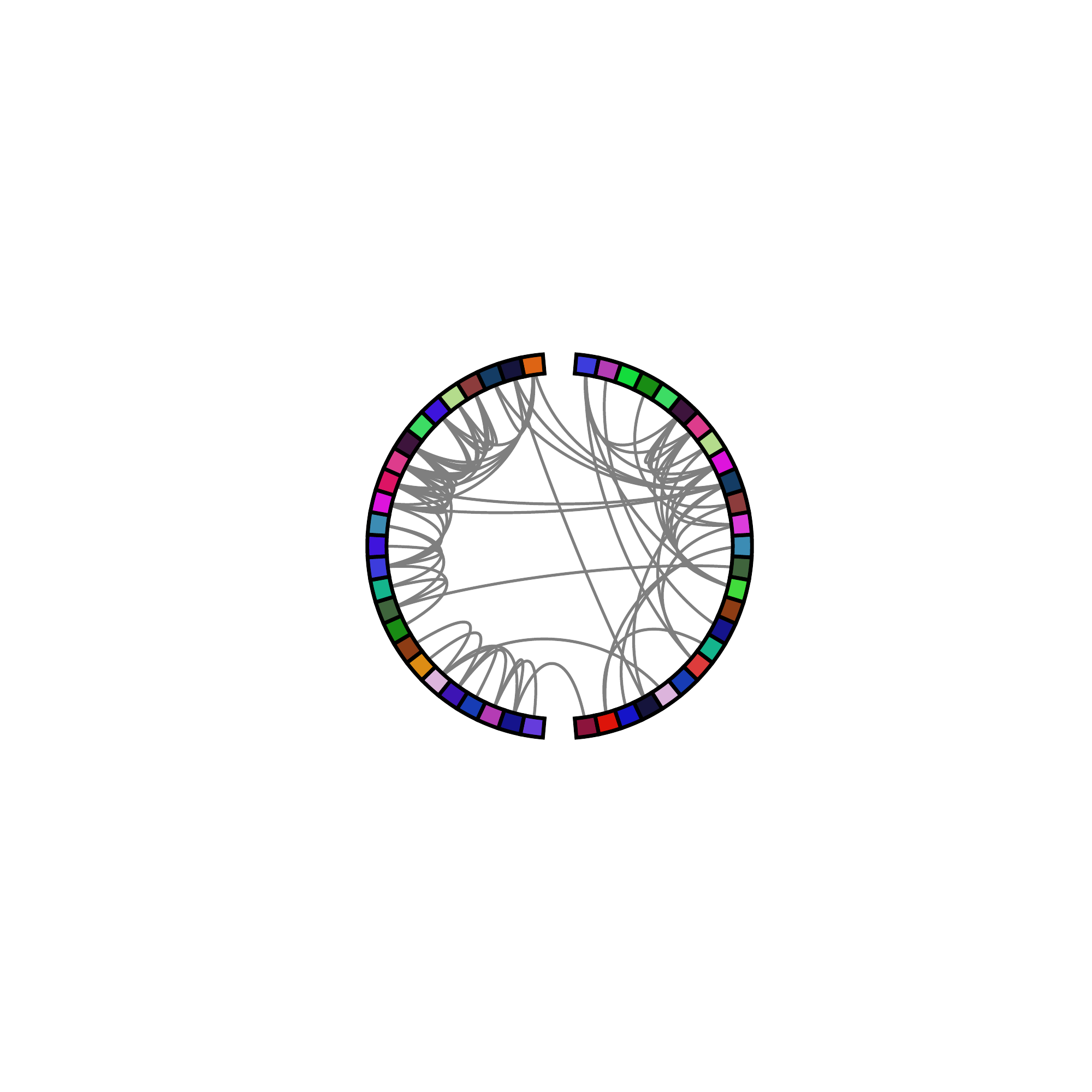} &
        \includegraphics[width=0.17\textwidth]{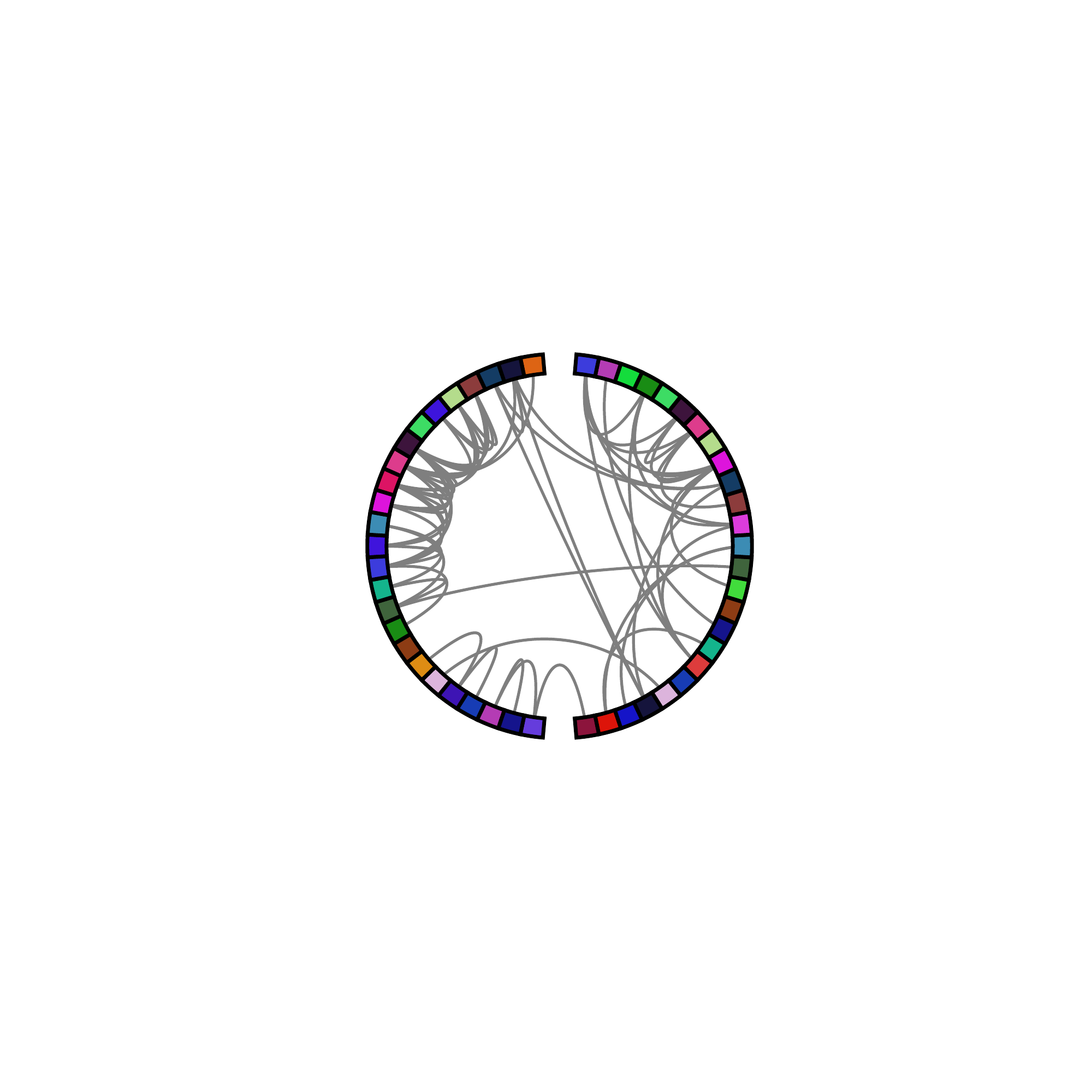} &
        \includegraphics[width=0.17\textwidth]{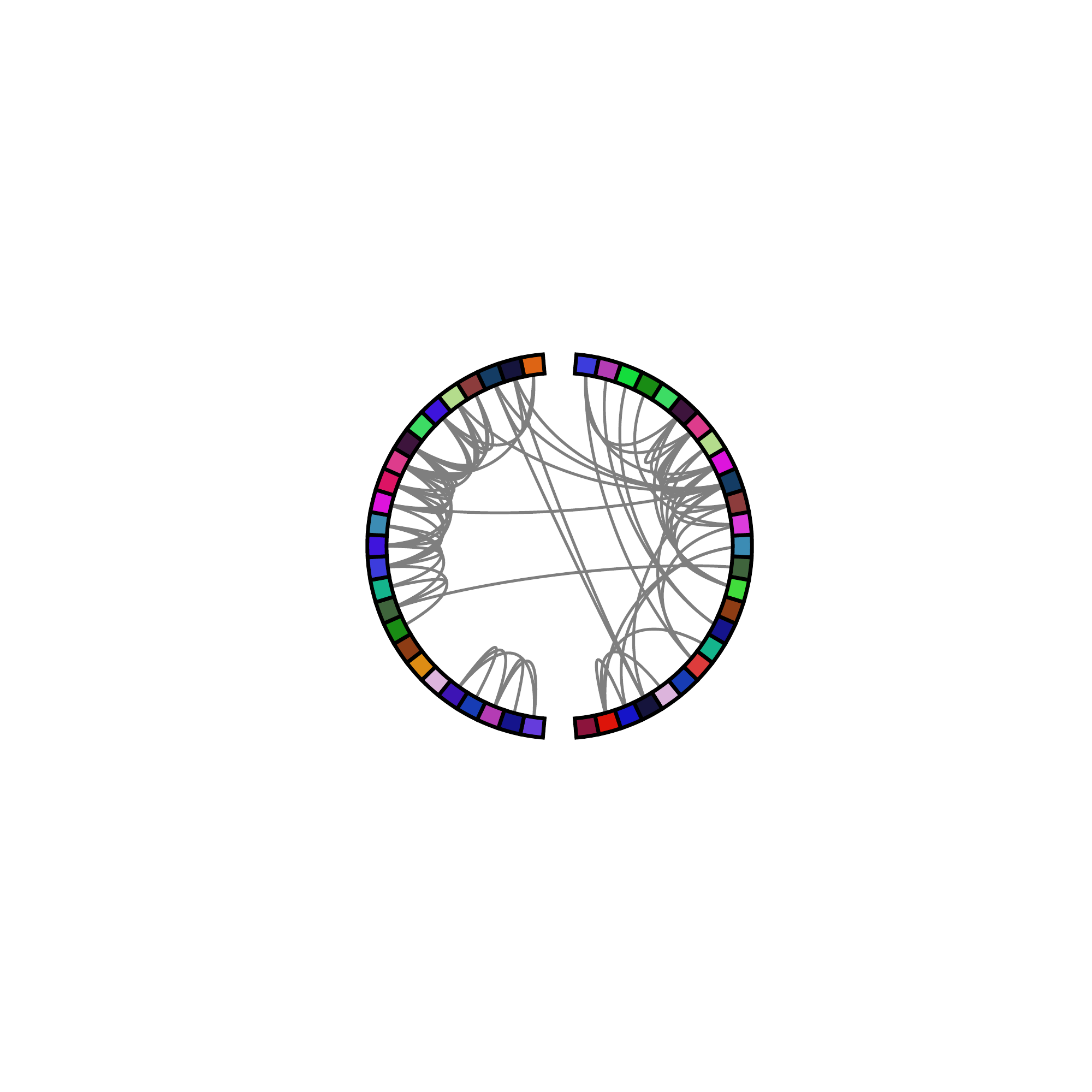} &
        \includegraphics[width=0.17\textwidth]{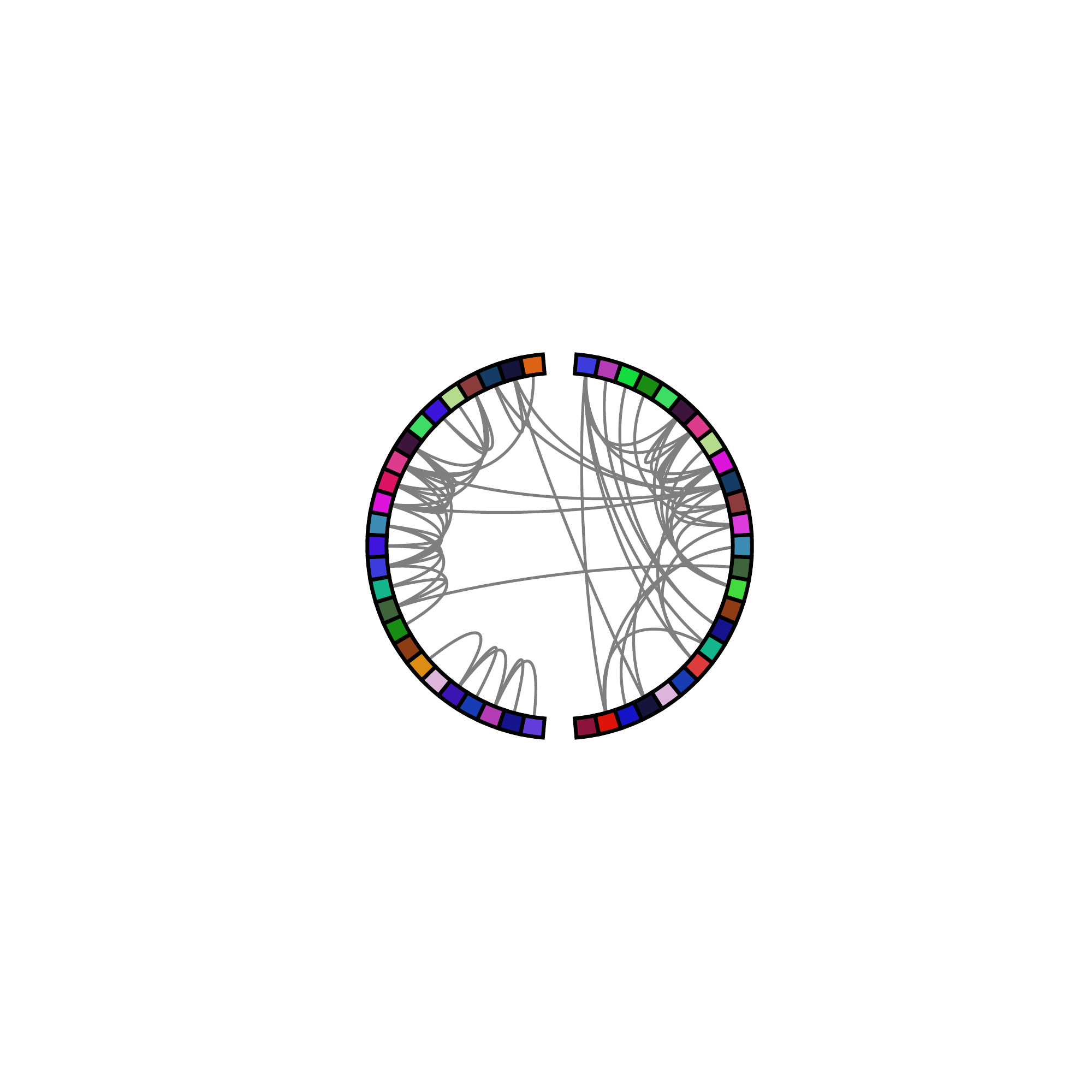} \\

        \includegraphics[width=0.17\textwidth]{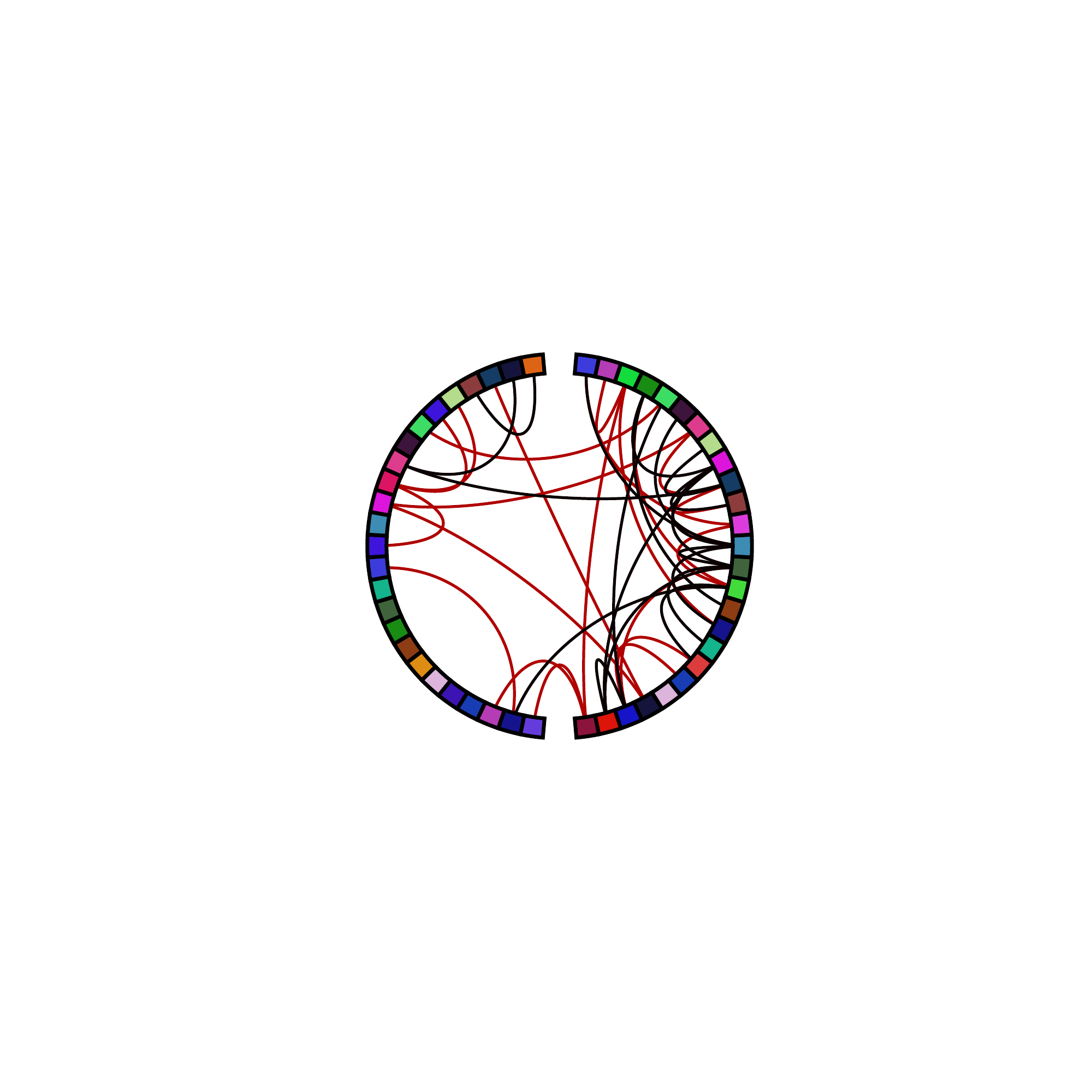} &
        \includegraphics[width=0.17\textwidth]{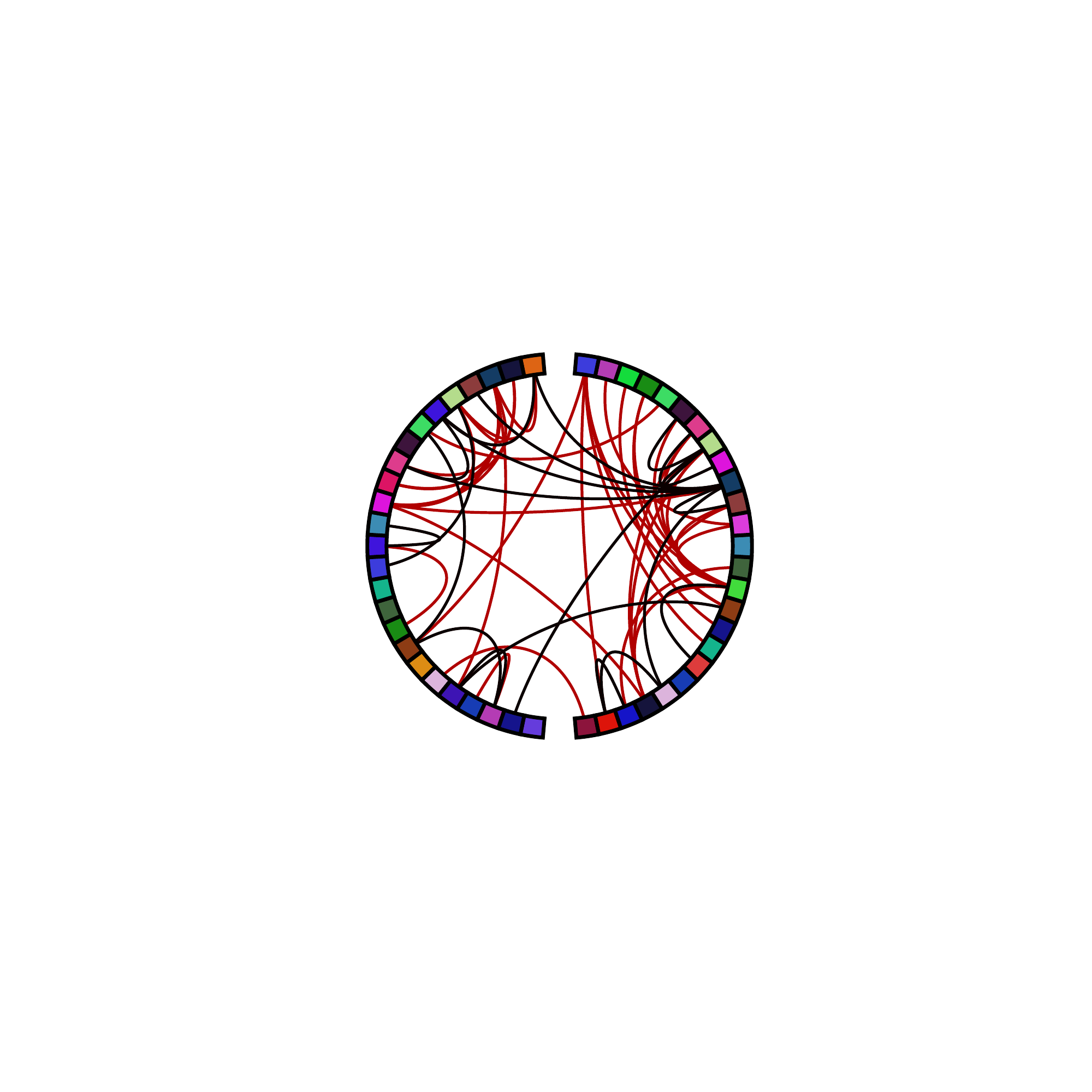} &
        \includegraphics[width=0.17\textwidth]{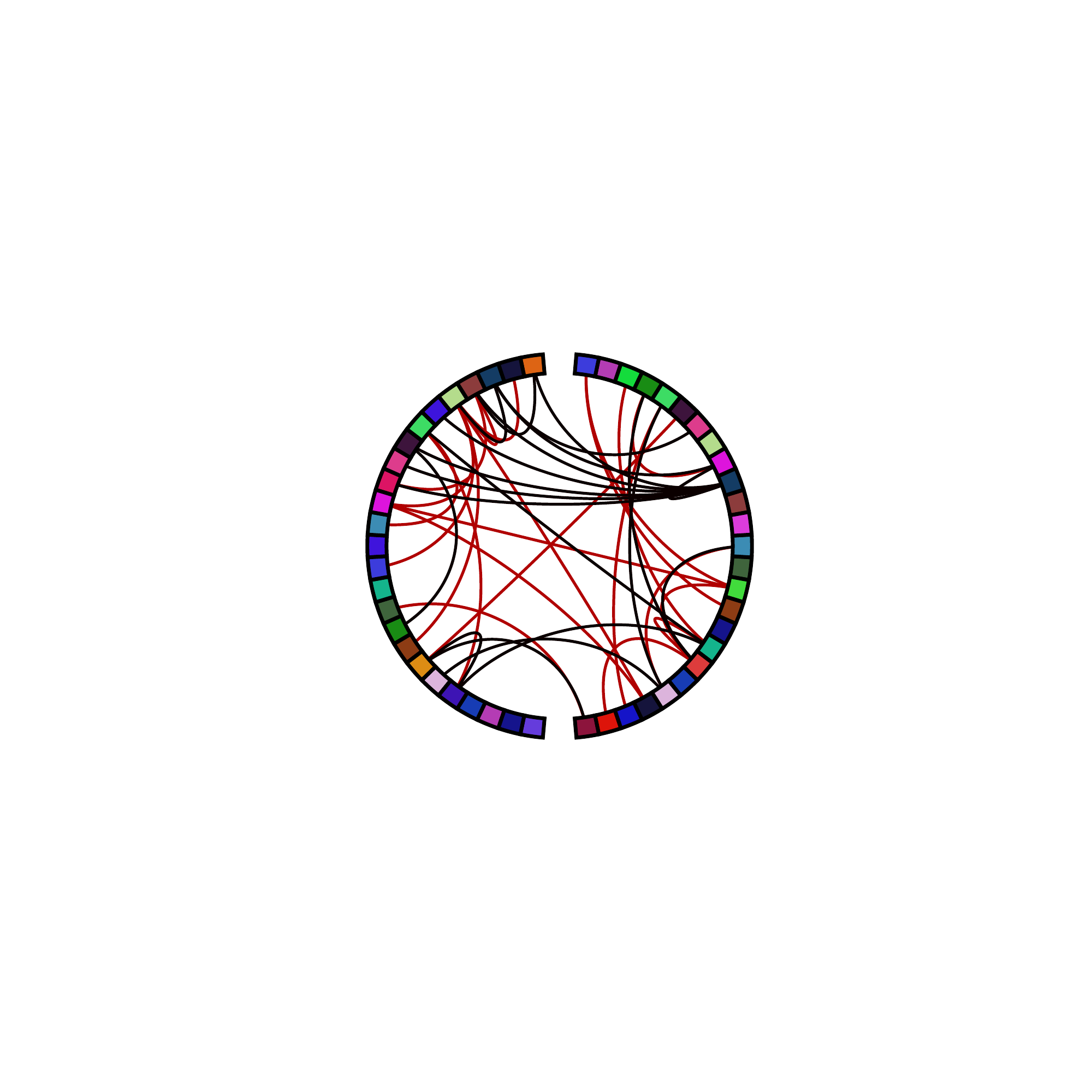} &
        \includegraphics[width=0.17\textwidth]{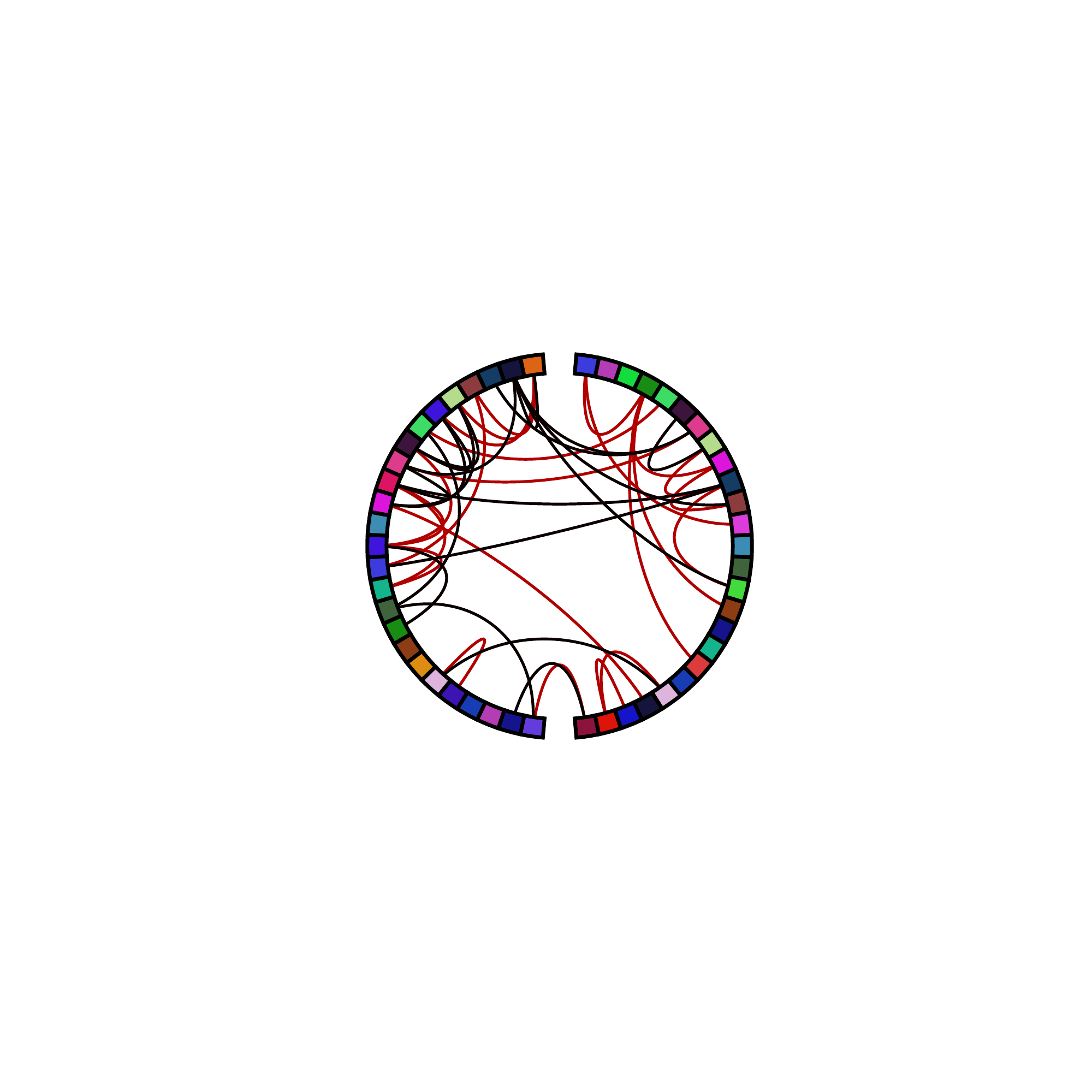} \\
    \end{tabular}
    \caption{Learned TGMs for different MEG conditions.  Each node on the periphery represents a brain region with location indicating anatomical location. \textbf{Top:}  Intersection of learned edges between switching and non-switching conditions. \textbf{Bottom:}  \textit{Black} edges indicating those in the non-switching condition but not in the switching and \textcolor{red}{\textit{red}} vice versa.} \label{fig:meg_intdiff}
    \vspace{-.2in}
\end{figure*}

Next we learn TGMs to capture the structure of underlying cortical dynamics
from magnetoencephalography (MEG) data collected from ten subjects who were
asked to perform a task while maintaining focus on an audio stream and then
again while switching focus~\cite{Larson:2014}.
Our goal is to discover differences in the underlying TGMs between the
non-switching and switching attention conditions.  Such differences provide
further understanding into the neural underpinnings of auditory selective
selection, an important constituent to communication.

The data were collected for each subject performing the experiment in
the \emph{switching} (S) and \emph{non-switching} (N) attention conditions.  For both S and N
conditions, each subject performed the task under an auditory condition of
\emph{high} (U) and \emph{low} (D) pitch, and spatial conditions of \emph{left} (L) and \emph{right} (R) attending.  For each
of the eight possible conditions, MEG recordings were collected resulting in a
150-dimensional time series of length 992 where each dimension corresponds to
a localized region of the brain.  We have between 17 and 30 trials
for each subject, resulting in about 200 replicate time series per condition.

Often with MEG data, many of the dimensions are dominated by noise due to
limited brain activity in that region.  We reduced the number of brain regions
we studied from 150 to 50 by only considering those with largest
variance.  In particular, for each trial we mean-centered all of the
time-series and computed the variance and retained the top 50 most volatile regions.

We computed the periodogram for each trial and averaged across trials within each condition,
resulting in eight periodograms.  
We ran our spectral TGS version of the FINCS algorithm on these periodograms for
100,000 iterations with fractional prior parameter $4 / N_c$, where $N_c$ is
the number of trials for condition $c \in \{\mbox{S, N}\} \times \{\mbox{U, D, L, R}\}$.
We also ran the algorithm for 1.7 million iterations and saw no
difference in the resulting graphs.

In Figure~\ref{fig:meg_intdiff}, we depict the intersections and
differences between the learned graphs for each experimental condition.  We see
in the top row that there are a lot of shared connections between the
switching and non-switching conditions for each auditory condition.  In the
bottom row, the differences between the switching and non-switching conditions
are depicted where red edges are those in the switching condition but not
the non-switching, and black edges are the reverse.  The difference plots
show that there seems to be substantial ``rewiring'' for many of the conditions
with many edges connecting frontal to back regions.  Interestingly, we again see consistencies in these rewirings across conditions.  Additionally, we reliably uncover local connections between adjacent brain regions across experimental conditions.  
Such observations provide guidance for developing experiments and methods to
discern the underlying mechanisms that give rise to these different structures.

\section{DISCUSSION}\label{sec:discussion} 
\vspace{-.2in}
We introduced a Bayesian approach to graphical model structure learning for time series.  In particular, we propose a prior---the \emph{hyper complex inverse Wishart} distribution---for the spectral density matrices in a Whittle likelihood approximation.  For decomposable graphs, this prior is conjugate and leads to a closed-form expression of the marginal likelihood of the time series given the graph, marginalizing the spectral density matrices across frequencies.  Being able to integrate out this large collection of complex matrices---one for each time point---is critical to developing a practical and scalable inference algorithm.  For this, exploiting the fact that our marginal likelihood is analogous to that for i.i.d. Gaussian graphical models \cite{jones:2005} but with a product over the number of Fourier frequencies, allows us to deploy straightforward modifications to existing MCMC and stochastic search algorithms. Our simulations show that when many time series are observed, our method recovers the correct graph. When a single time series is observed, we proposed a method to increase robustness of our graph estimation using a piecewise constant prior. 
Our results on the stock and MEG datasets demonstrated our ability to discover intuitive and interpretable structure in these datasets, importantly leveraging the temporal dependencies.

Extensions to non-decomposable graphs are possible using the i.i.d. graph approaches in both \cite{Roverato:2002} and \cite{Moghaddam:2009}. A Laplace approximation to the marginal likelihood for non-decomposable graphs is proposed in \cite{Moghaddam:2009}, which we could similarly utilize to approximate the frequency-specific marginal at each term in Equation~\eqref{eq:marg}. Parallelizing the Laplace approximation computation across frequencies would lead to a scalable method for inference in non-decomposable time series graphs.

\vspace{-.1in}
{\small
{\bf Acknowledgements:} This work was supported by DARPA Grant FA9550-12-1-0406 negotiated by AFOSR, ONR Grant N00014-10-1-0746, NSF CAREER Award IIS-1350133, and AFOSR Grant FA9550-12-1-0453.
}


{\small
\bibliography{timegraphs}
\bibliographystyle{unsrt}
}
\nopagebreak
\setcounter{section}{0}
\setcounter{figure}{0}
\setcounter{table}{0}
\setcounter{equation}{0}
\setcounter{footnote}{0}

\begin{center}
\twocolumn[
\centering
\textbf{\Large{Bayesian Structure Learning for Stationary Time Series:\\ Supplementary Material}}
\vspace{.5in}]
\end{center}

\vspace{3in}

\maketitle
\begin{abstract} In this supplement we provide more background
    on spectral analysis of time series and the complex normal and complex inverse Wishart distributions.  The hyper
    complex inverse Wishart is then introduced in more detail and its marginal
    likelihood is derived. We also provide more detail about the periodogram
    smoothing plug in method referenced in Section~4 of the main text. 
    Finally, we provide details about the global stock index data set and compare the underlying conditional independence graphs learned by our method with that of Songsiri et al.~\cite{Songsri:2011}. We also provide supplemental simulations showcasing the predictive performance of our approach in the frequency domain. 

\end{abstract}


\section{Spectral Analysis Background}

Spectral analysis is an approach to
analyzing stationary time series where the main object of interest is the
\emph{spectral density}. For many applications, the spectral density is often a
more informative object about the underlying physical process than
autoregressive coefficients or the lagged autocovariance. In particular, it
allows one to read off which frequency components are prominent in a time
series, and usually certain frequency bands have scientific relevance. The
theoretical justification for caring about the spectral density arises from the
classic \emph{spectral representation theorem} \cite{Brockwell:1991}.
Informally, this theorem states
that under some conditions, any stationary vector stochastic process $x(t) \in
\mathbb{R}^{p}$ can be written as
\begin{equation}
x(t) = \int_{-\pi}^{\pi} e^{-i t \lambda} dZ(\lambda)
\end{equation}
where $d Z(\lambda) \in \mathbb{C}^{p}$ is an \emph{orthogonal increments
process} such that $E (Z(\lambda) Z(\lambda)^*) = S(\lambda)$, where
$S(\lambda)$ is the spectral density matrix and $E (Z(\lambda) Z(\lambda')^*) =
0$ for $\lambda \neq \lambda'$. Intuitively, this theorem states that the
amplitudes of certain frequencies in a stationary process are independent
across frequencies and the within frequency covariance is given by the spectral density matrix. As noted in the main text, the spectral density matrix also arises as the Fourier transformation of the lagged auto covariance matrices of the process
\begin{equation}
S(\lambda) = \sum_{h = - \infty}^{\infty} \Gamma(h) e^{-i \lambda h}
\end{equation}
where $\Gamma(h) = E(x(t) x(t + h)^T)$. The statistical task then becomes estimation of the spectral density matrix $S(\lambda)$ from a finite observed time series $x = [x(1),...,x(T)]$. The most basic estimator is given by the periodogram, defined as:
\begin{equation}
P_k = d_k d_k^*
\end{equation}
where $d_k$ is the DFT of the observed time series at Fourier frequency
$\lambda_k$, (Eq. (7) of the main text). While this estimator is asymptotically
unbiased, it is not consistent since its variance does not go to zero. Instead,
techniques that smooth across nearby frequencies provide consistent estimators
of the spectral density, and are commonly used in practice \cite{Percival:1993}.  Finally, the
periodogram estimates at different frequencies, $P_k$ and $P_k'$, are
asymptotically uncorrelated, providing some intuition as to why the Whittle
approximation decomposes into independent terms for each frequency~\cite{Brillinger:2001}. More details on the spectral approach to analyzing stationary time series is provided in Brillinger, 2001~\cite{Brillinger:2001}.

\section{Smoothing the Periodogram for a Single Time Series via the Plug in Method} \label{Sec:smoothing}

In this section we provide more details on the plug in method for smoothing the
periodogram obtained from a single realization of a multivariate time series
mentioned in Sec. 4 of the main text. First we provide some background on
classical frequentist approaches to smoothing the periodogram to obtain
consistent estimators \ of the spectral density (as $T$ increases).

When the spectral density itself is the primary object of interest, a common frequentist method is to smooth the periodogram to obtain a consistent estimator of the spectral density:
\begin{equation}
\hat{S}(\lambda_k) = \sum_{|j| < m} W_T(j) P_{k + j}
\end{equation}
where  $P_{k}$ is the periodogram at frequency $\lambda_k$ as introduced in the
main text and $W_T(j) \geq 0$, $\sum_{|j| < m} W_T(j) = 1$ are some smoothing
weights for a length $T$ series and $m$ is the smoothing window. This approach
was used in the frequentist graph estimation frameworks in \cite{Jung:2014,
Bach:2004, Dahlhaus:2000}. To ensure consistency as $T \to \infty$ we must have
$m \to \infty$, $\frac{m}{n} \to 0$,   and $\sum_{|j| < m} W_T(j)^2 \to 0$
\cite{Brockwell:1991}. The asymptotic variance of $\hat{S}_k$ scales as
$\sum_{|j| \leq m} W_T^2(j)$, implying that the asymptotic effective sample
size for a smoothed estimate of this form is $(\sum_{|j| \leq m}
W_T^2(j))^{-1}$ \cite{Brockwell:1991}. The Daniell smoother corresponds to
taking $W_T(j) = \frac{1}{2m+1}$ and has an intuitive (effective) sample size of $2m + 1$, the size of the smoothing window. Intuitively, this holds asymptotically since as $T \to \infty$ the sample periodograms become independent at different frequencies implying a sample size of $2m + 1$, the number of (asymptotically) independent samples. 

Inspired by the use of this smoothing technique in previous TGM procedures
$\cite{Jung:2014, Bach:2004, Dahlhaus:2000}$ we develop a similar procedure
tailored to our objective function in Eq. \eqref{eq:marg}. We plug in a
smoothed estimate of the spectral density matrix, scaled by the asymptotic
effective degrees of freedom, for the priodogram, $P_k$, in Eq. \eqref{eq:marg}. Specifically, we set $W^*_k = W_k + (\sum_{|j| \leq m} W_T^2(j))^{-1} \hat{S}_k$.  The degrees of freedom parameter $\delta^*_k$ is similarly updated by adding the effective sample size of the smoother to the prior degrees of freedom: $\delta^*_k = \delta_k +  (\sum_{|j| \leq m} W_T^2(j))^{-1}$. If we use the Daniell smoother outlined above the updates become $W^*_k = W_k + \sum_{|j| \leq m} P_{k + j}$ and $\delta^* = \delta_k + 2m + 1$. In practice we set $m = \lfloor \frac{\sqrt{T}}{2} \rfloor$ to ensure that the conditions for consistency of $\hat{S}_k$ are met.

\section{The Complex Normal and Complex Inverse Wishart Distributions}

The complex normal distribution is a generalization of the multivariate normal
distribution to the complex domain. Let $Z \in \mathbb{C}^{p}$ be a complex
random variable. $Z$ is distributed as a complex normal distribution,
$\mathcal{N}_c(0, \Sigma)$, with zero mean and complex Hermitian positive definite covariance matrix $\Sigma \in \mathbb{C}^{p \times p}$ if it has density given by
\begin{equation} \label{eq:complexnormal}
p(z) = \frac{1}{\pi^p |\Sigma|} e^{-z^* \Sigma^{-1} z},
\end{equation}
where $z^{*} = \bar{z}^T$ denotes the conjugate transpose of $z$. If $Z \sim \mathcal{N}_c(0, \Sigma)$ then the distribution over $Z$ can be represented equivalently as a joint distribution over the real and imaginary elements of $Z = X + i Y$, $X, Y \in \mathbb{R}^{p}$
\begin{equation}
\left[\begin{array}{c}
X \\
Y
\end{array} \right]  \sim N(0, \left[\begin{array}{cc}
\Real \Sigma& - \Imag \Sigma\\
\Imag \Sigma & \Real \Sigma \\
\end{array} \right] ,
\end{equation}
where $\Real \Sigma$ and $\Imag \Sigma$ indicate the real and imaginary
components of $\Sigma$, respectively. Thus we see that the real and imaginary
components are independent iff $\Imag \Sigma = 0$.  As in the non-complex case,
the marginal likelihood of $X_A$ for some subset of nodes $A \subseteq
\{1,\ldots,p\}$, is given by $X_A \sim \mathcal{N}_c(0,\Sigma_{A})$, where $\Sigma_A$ is the matrix formed by selecting the rows and columns of $\Sigma$ in $A$. 


The conjugate prior distribution for $\Sigma$ is given by the complex inverse
Wishart, $\Sigma \sim IW_c(\delta,W)$, with degrees of freedom parameter
$\delta > 0$ and centering matrix $W \in \mathbb{C}^{p \times p} $, Hermitian positive definite. Its density is given by
\begin{equation}
p(\Sigma|W,\delta) = B(W,\delta) |\Sigma|^{-(\delta + 2 p)} e^{- \text{tr} W \Sigma^{-1}}
 \end{equation}
with normalization constant
\begin{equation*}
    B(W,\delta) = \frac{|W|^{\delta + p}}{\pi^{\frac{p(p - 1)}{2}} \prod_{j
    = 1}^{p} (\delta + p - j)!}.
\end{equation*}
Note that we have used an alternative parameterization of the inverse Wishart
distribution commonly used in the graphical modeling literature
\cite{Giudici:1999}. The marginal distribution of $\Sigma_A$ where $A \subseteq
\{1,\ldots,p\}$ is given by $\Sigma_{A} \sim IW_c(\delta,W_{A})$.

\section{Marginal Likelihood for the Hyper Complex Inverse Wishart}

We define the hyper-complex inverse Wishart distribution for a graph $G = \{V,E\}$ in the main paper as the restriction of the complex inverse Wishart distribution to $\Sigma \in \mathbb{C}^{p \times p}$ with a zero pattern in $\Sigma^{-1}$ specified by $G$. Its density is given by:
\begin{align} \label{eq:hyper}
p(\Sigma|\delta,W,G ) = {\BF 1}_{\Sigma \in M^{+}(G)} h(W,\delta,G)  |\Sigma|^{-(\delta + 2 p)} e^{- \text{tr} W \Sigma^{-1}}
\end{align}
where $h(W,\delta,G)$ is a normalization constant and
$M^{+}(G)$ is the set of positive definite matrices with zeros in their inverse that obey the conditional independence properties of $G$.

Due to the fact that the complex inverse Wishart distribution is conjugate to
the complex normal distribution for an unrestricted $\Sigma$, by Proposition
5.1 in \cite{Dawid:1993} it follows that the hyper complex inverse Wishart
distribution is a strong hyper-Markov distribution. It follows that for decomposable $G$ the complex hyper inverse Wishart density can be written in terms of the cliques, $\mathcal{C}$, and separators, $\mathcal{S}$, of $G$:
\begin{equation}
p(\Sigma|\delta,W,G ) =  {\BF 1}_{\Sigma \in M^{+}(G)} \frac{\prod_{C \in \mathcal{C}} p(\Sigma_C|W_C,\delta)}{\prod_{S \in \mathcal{S}} p(\Sigma_S|W_S,\delta)}
\end{equation}
where $p(\Sigma_C|W_C,\delta)$ is the unrestricted complex inverse Wishart density for $\Sigma_C$.  This decomposition implies that the normalization constant for Equation \eqref{eq:hyper} is also given by the ratio of complex inverse Wishart normalization constants for cliques and separators
\begin{equation}
h(W,\delta,G) = \frac{\prod_{C \in \mathcal{C}} B(W_C, \delta)}{\prod_{S \in \mathcal{S}} B(W_S,\delta)}.
\end{equation}
If $Z_1,...Z_N \stackrel{i.i.d.}{\sim} \mathcal{N}_c(0,\Sigma)$, then the joint
distribution of $Z_1,\ldots, Z_N$, and $\Sigma$ can be written as:
\begin{align}
p(z_1,\ldots,z_N, \Sigma|G, W, \delta) &\propto \\
 {\BF 1}_{\Sigma \in M^{+}(G)} \frac{h(W,\delta,G)}{\pi^{N p}}  |\Sigma|^{ -(\delta + N + 2p)}   & e^{- \text{tr} (W + \sum_{i = 1}^{N} z_i z_i^*) \Sigma^{-1}} .
\end{align}
We note that the part dependent on $\Sigma$ is the kernel for a $HIW_c(W +
\sum_{i = 1}^{N} z_i z_i^*, \delta + N, G)$ distribution, it follows that the
marginal distribution of $Z_1,\ldots,Z_n|G, W, \delta$ is given by the ratio of
prior and posterior normalization constants of the complex hyper inverse
Wishart distribution times a likelihood constant:
\begin{equation}
p(z_1, \ldots, z_n|G, W, \delta) = \frac{h(W,\delta,G)}{\pi^{Np} h(W + \sum_{i = 1}^{N} z_i z_i^*,\delta + N,G)}.
\end{equation}

\subsection{Marginal Whittle Likelihood}

The model in the main text places independent $HIW_c(W_k, \delta_k, G)$ priors
on each spectral density matrix in the Whittle likelihood, $S_k \sim HIW_c(W_k,
\delta_k, G)$  $\forall k \in [T - 1]$.  Applying the above marginal likelihood
result to each frequency component in the Whittle approximation shows that the
marginal likelihood of the data given a graph, a set of centering matrices
,$W_0,\ldots, W_{T - 1}$, and degrees of freedom, $\delta_0, \ldots
\delta_{T-1}$, for each frequency can be approximated by a product of the normalization constants across frequencies:
\begin{equation}\label{eq:marg}
p(\BF{X}_{1:N}|G) \approx \pi^{-N T p} \prod_{k = 0}^{T - 1}  \frac{h(W_k,\delta_k,G)}{h(W^*_k, \delta^*_k,G)}.
\end{equation}
where $W^*_k = W_k + P_k$ and $\delta^*_k = \delta_k + N$. Indeed, this derivation shows that our prior specification for spectral density matrices is conjugate to the entire Whittle likelihood.

\section{Prediction Simulations}
To further validate our approach we analyze predictive performance in the frequency domain 
on simulated VAR(1) data as described in Section 6 of the main text. We set the dimension to $p = 10$ and time series length to $T = 2500$. We split
the simulated series in half forming a training set and test set and then learn a graph and
smoothed periodogram on the test set of the time series. We use the
Whittle marginal likelihood presented in Eq.~\eqref{eq:marg} to compare
predictions between four graphs: the learned graph in the frequency domain, $\hat{G}_{\text{spectral}}$, the learned iid graph that treats the time series as independent observations, $\hat{G}_{\text{iid}}$ , the full graph, $G_{\text{full}}$,  with all edges included, and the empty
graph, $G_{\text{empty}}$, with no edges. For prediction, the prior centering
matrix, $W_k$, is given by the Daniell smoothed periodogram on the training data
with its respective degrees of freedom, and $W^{*}_k$ is given by $W^{*}_k =
W_k + P^{\text{test}}_k$, where $P^{\text{test}}_k$ is the periodogram on the
test data. Results are displayed in Table 1. We see that the learned spectral graph does significantly better than the other graphs at prediction.

\begin{table*}
\centering
\begin{tabular}{|c | c | c | c| c |}
\hline
 & $\hat{G}_{\text{spectral}}$ & $\hat{G}_{\text{iid}}$ & $G_{\text{empty}}$ & $G_{\text{full}}$ \\
 \hline
 Loglik. & -81623 $\pm$ 205 & -87712 $\pm$ 216 &  -90040 $\pm$ 211 & -103964 $\pm$ 304 \\
\hline
\end{tabular}
\caption{Predictive marginal log-likelihood for simulated VAR data in the
frequency domain on a held out test set using the marginal Whittle likelihood
in Eq.~\eqref{eq:marg} under four different graphs ($\pm$ indicates 1 standard
error across simulation replicates).}
\label{predtabsim}
\end{table*}

\section{Global Stock Indices}

\subsection{Learned Graph Comparison with Previous Methods}
For comparison with the method of Songsiri et al. \cite{Songsri:2011}, we
provide the CIG graphs learned on the international stock data set using both
their autoregressive method and our nonparametric Bayesian method in Figure
\ref{graph_comp}. Further details on the meaning of the edge weights can be
found in \cite{Songsri:2011}.  Notice that both graphs capture similar
structure, for instance the connections between the US, Canada, Australia, and
Japan.  Additionally, both graphs contain tight clusters containing European
countries.

\begin{figure*}[t!] \label{graph_comp}
\includegraphics[width=.4\textwidth]{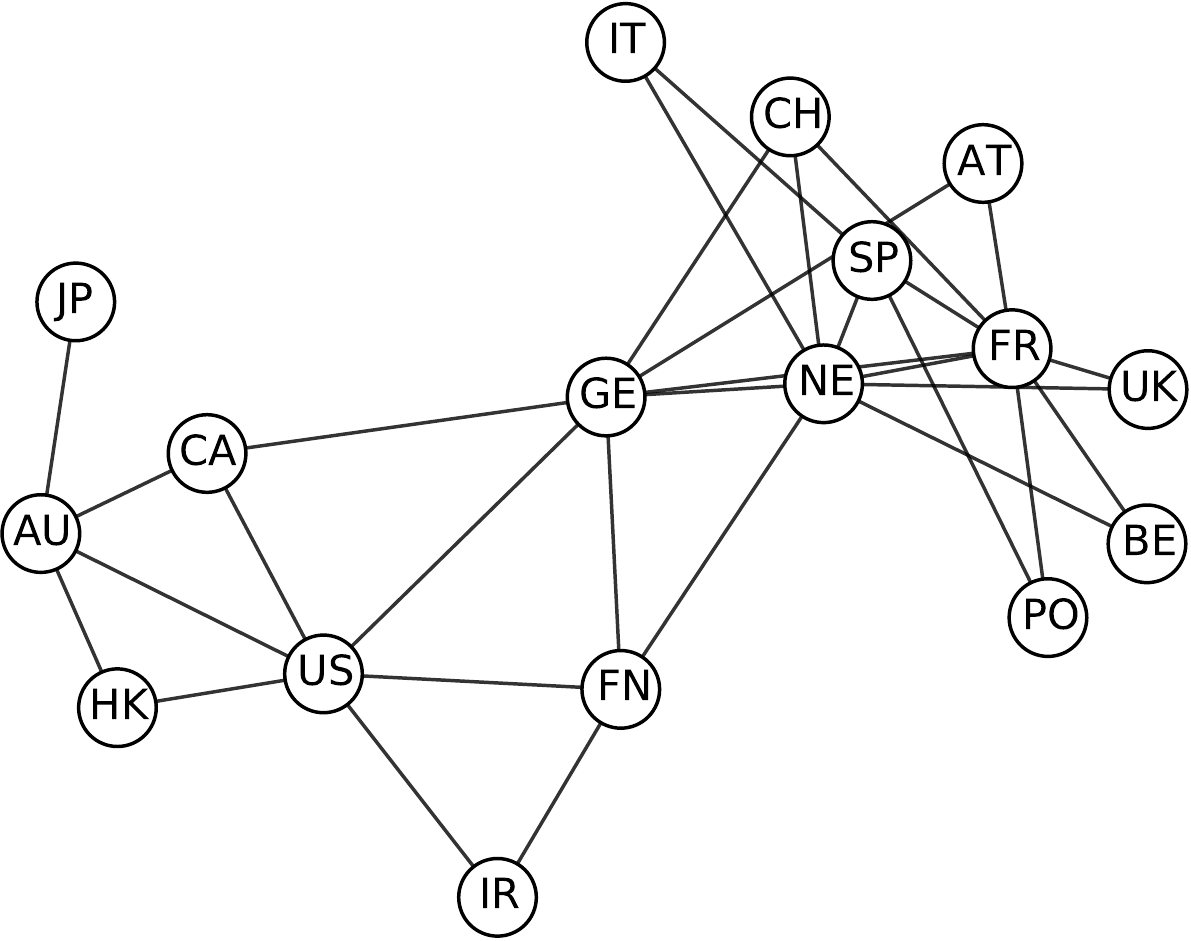} 
\includegraphics[width=.68\textwidth]{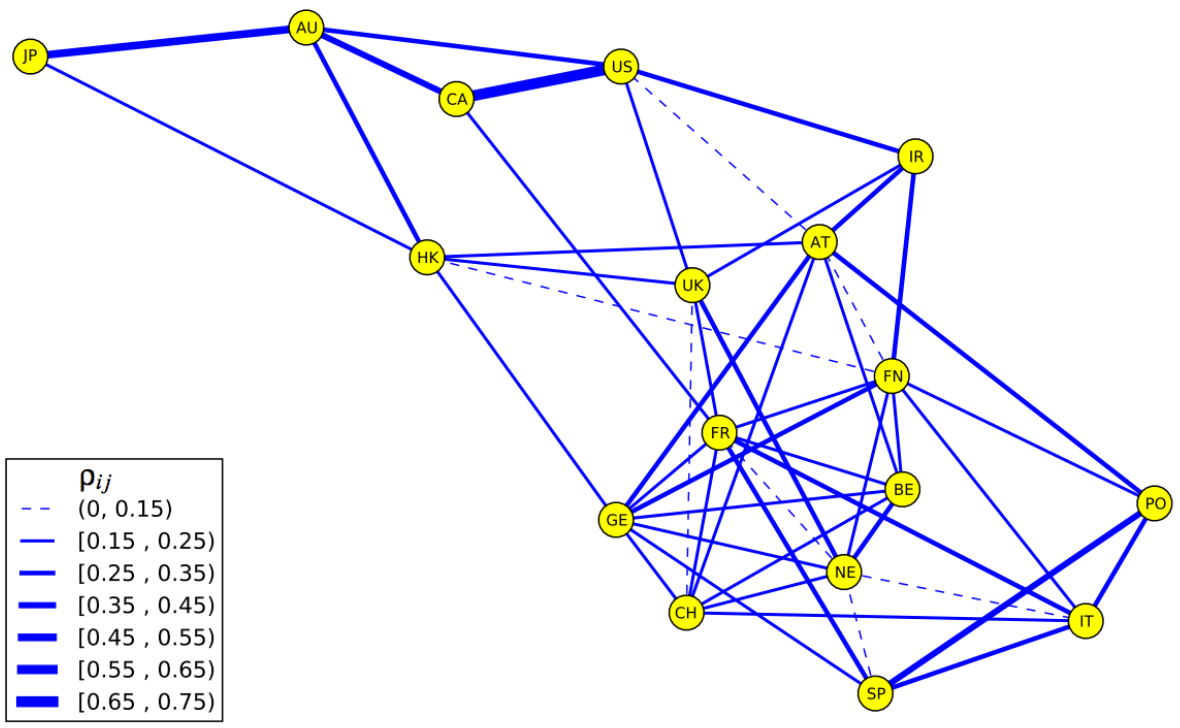}
\caption{CIG graph learned on international stock data using (\emph{left})
method presented in this paper and (\emph{right}) the autoregressive method
presented in Songsiri et al.~\cite{Songsri:2011} (Figure taken directly from
\cite{Songsri:2011}).}
\label{graph_comp}
\end{figure*}

\subsection{Stock Data}
In Table~\ref{tab:stocks}, we list the stock indices that were used in the main
paper.  The data was downloaded from \texttt{globalfinancialdata.com} for the
dates June 3, 1997 to June 30, 1999.

\begin{table*}
    \centering
    \caption{Stock index information.}
    \begin{tabular}{cccc}
    \label{tab:stocks}
        Index Name & Ticker & Country & Country Code \\
        \hline
    Amsterdam Exchange Index & AEX & Netherlands & NE \\
        All Ordinary Composite & AORD & Australia & AU \\
        Austrian Traded Index & ATX & Austria & AT \\
        BEL 20 & BFX & Belgium & BE \\
        CAC 40 & FCHI & France & FR \\
        FTSEMIB & FTMIB & Italy & IT \\
        FTSE 100 & FTSE & United Kingdom & UK \\
        DAX 30 & GDAX & Germany & GE \\
        Toronto Stock Exchange 300 & GSPTSE & Canada & CA \\
        Hang Seng Composite & HSI & Hong Kong & HK \\
        IBEX 35 & IBEX & Spain & SP \\
        Irish Stock Exchange Index & ISEQ & Ireland & IR \\
        Nikkei 225 & N225 & Japan & JP \\
        OMX Helsinki 25 & OMXH25 & Finland & FN \\
        Portugal Stock Index & PSI20 & Portugal & PO \\
        S\&P 500 & SPX & United States & US \\
        Swiss Market Index & SSMI & Switzerland & CH \\
    \end{tabular}
\end{table*}

\end{document}